\documentclass[a4paper,11pt]{article} 

\usepackage[english]{babel}
\usepackage[utf8]{inputenc}
\usepackage[T1]{fontenc}
\usepackage{graphicx}
\usepackage{wrapfig}
\usepackage[toc,page]{appendix}
\usepackage{amsmath}
\usepackage[labelfont=bf]{caption}
\usepackage{enumerate}
\usepackage{enumitem}
\usepackage{siunitx}
\usepackage{subfigure}
\usepackage{fancyhdr}
\usepackage{courier}
\usepackage{array, booktabs}
\usepackage{amsthm}
\usepackage{amsfonts}
\usepackage[utf8]{inputenc}
\usepackage[T1]{fontenc}
\usepackage{calligra}
\usepackage{makeidx}
\usepackage{amssymb}
\usepackage{frontespizio}
\usepackage{makecell}
\usepackage{tikz}
\usetikzlibrary{fit}
\usepackage[hyphenbreaks]{breakurl}
\usetikzlibrary{automata,arrows,positioning,calc}
\usetikzlibrary{decorations.markings}
\usetikzlibrary{matrix}
\usepackage[bookmarksdepth=subsection,plainpages=false]{hyperref}

\newcommand{\be}{\begin{eqnarray}}
\newcommand{\ee}{\end{eqnarray}}

\usepackage[a4paper, inner=1.9cm, outer=1.9cm, top=2cm, 
bottom=2cm, bindingoffset=0cm]{geometry}

\begin{document} 

\title{\huge{Antimalarial Artefenomel Inhibits Human SARS-CoV-2 Replication in Cells while Suppressing the Receptor ACE2
}} 
\date{}
\maketitle

\vspace{-1.8cm}
\begin{center}
\noindent Tania Massignan\textsuperscript{1*}, Alberto Boldrini\textsuperscript{1*}, Luca Terruzzi\textsuperscript{1*}, Giovanni Spagnolli\textsuperscript{2,3*}, Andrea Astolfi\textsuperscript{1,4*},  \\  Valerio Bonaldo\textsuperscript{2,3}, Francesca Pischedda\textsuperscript{2}, Massimo Pizzato\textsuperscript{2}, Graziano Lolli\textsuperscript{2}, Maria Letizia Barreca\textsuperscript{4}, Emiliano Biasini\textsuperscript{2,3,\#}, Pietro Faccioli\textsuperscript{5,6,\#} \& Lidia Pieri\textsuperscript{1\#}
\\
\vspace{0.3cm}
\noindent \textsuperscript{1} Sibylla Biotech SRL, Verona VR, Italy \\
\noindent \textsuperscript{2} Department of Cellular, Computational and Integrative Biology (CIBIO) and \textsuperscript{3} Dulbecco Telethon \\  \indent{} Institute, University of Trento, Trento TN, Italy\\
\noindent \textsuperscript{4} Department of Pharmaceutical Sciences, University of Perugia, Perugia PG, Italy\\
\noindent \textsuperscript{5} Department of Physics, University of Trento and \textsuperscript{6} INFN-TIFPA, Trento, Italy\\
\vspace{0.3cm}
*Authors Contributed Equally \\
\textsuperscript{\#}Correspondence: emiliano.biasini@unitn.it; pietro.faccioli@unitn.it; lidia.pieri@sibyllabiotech.it
\end{center}
\section*{Abstract}
The steep climbing of victims caused by the new coronavirus disease 2019 (COVID-19) throughout the planet is sparking an unprecedented effort to identify effective therapeutic regimens to tackle the pandemic. The SARS-CoV-2 virus is known to gain entry into various cell types through the binding of one of its surface proteins (spike) to the host Angiotensin-Converting Enzyme 2 (ACE2). Thus, the spike-ACE2 interaction represents a major target for vaccines and antiviral drugs. A novel method has been recently described by some of the authors to pharmacologically downregulate the expression of target proteins at the post-translational level. This technology builds on computational advancements in the simulation of folding mechanisms to rationally block protein expression by targeting folding intermediates, hence hampering the folding process. Here, we report the all-atom simulations of the entire sequence of events underlying the folding pathway of ACE2. Our data revealed the existence of a folding intermediate showing two druggable pockets hidden in the native conformation. Both pockets were targeted by a virtual screening repurposing campaign aimed at quickly identifying drugs capable to decrease the expression of ACE2. We identified four compounds (the atypical antipsychotic Ziprasidone, the antihistamine Buclizine, the antiviral Beclabuvir and the antimalaric Artefenomel) as capable of lowering ACE2 expression in Vero cells in a dose-dependent fashion. All these molecules were found to inhibit the entry into cells of a pseudotyped retrovirus exposing the SARS-CoV-2 spike protein. Importantly, the antiviral activity has been tested against live SARS-CoV-2 (MEX-BC2/2020 strain). Our results indicate that one of the selected drugs (Artefenomel) could completely prevent cytopathic effects induced by the presence of virus, thus showing a promising antiviral activity against SARS-CoV-2. Ongoing studies are further evaluating the possibility of repurposing these drugs for the treatment of COVID-19.

\section*{Introduction}
Severe acute respiratory syndrome coronavirus 2 (SARS-CoV-2) is the causative agent of the viral pneumonia outbreak named Coronavirus Disease 2019 (COVID-19), a pandemic started at the end of 2019 in the Hubei province of China and later spread to the rest of the planet\textsuperscript{\tiny{1}}. The disease shows unprecedented transmission, morbidity and mortality rates\textsuperscript{\tiny{2,3}}. In the absence of a vaccine,  effective therapeutics to control the disease are in urgent need. SARS-CoV-2 belongs to the family of Coronaviridae, positive-sense, single-stranded RNA viruses of humans and animals causing pathologies ranging from a common cold to severe respiratory diseases, such as the Severe Acute Respiratory Syndrome (SARS) and the Middle East Respiratory Syndrome (MERS)\textsuperscript{\tiny{4}}. Similarly to other coronaviruses, SARS-CoV-2 has four main structural proteins, known as the nucleocapsid, membrane, envelope and spike\textsuperscript{\tiny{5}}. The latter promotes the binding of the virus to the extracellular domain of the angiotensin-converting enzyme 2 (ACE2)\textsuperscript{\tiny{6-8}}. Subsequently, another protein of the host, the transmembrane protease serine 2 (TMPRSS2), cleaves the spike protein, exposing a fusion peptide that promotes virus entry into the cell\textsuperscript{\tiny{9}}. ACE2 belongs to the protein family of angiotensin-converting enzymes\textsuperscript{\tiny{10}}. The protein is reported to be expressed in cells of the kidney, intestine, arteries, heart, and lung, although discrepancies about its expression profiles in different organs recently emerged\textsuperscript{\tiny{11,12}}. The main function of ACE2 is connected to its ability to lower blood pressure by catalyzing the conversion of angiotensin II into the vasodilator angiotensin\textsuperscript{\tiny{13}}. The protein is a promising drug target for treating cardiovascular diseases\textsuperscript{\tiny{14,15}}. 
Consistent with the mode of entry of SARS-CoV-2 into human cells, hampering the interaction of spike with ACE2 should effectively counteract virus replication. This objective could be achieved by different strategies, including antibodies against spike, compounds interfering with the spike-ACE2 interaction, or molecules capable of lowering the exposure of ACE2 at the cell surface\textsuperscript{\tiny{16}}.
The latter strategy could overcome possible evolutionary variations of the virus conferring resistance to treatments directed against viral proteins. However, the function of ACE2 on vascular regulation has been shown to play protective roles against virus-induced lung injury, and the complete silencing of the protein would likely result in severe secondary effects\textsuperscript{\tiny{6,17}}. For this reason, any pharmacological approach aimed at targeting ACE2 should be able to down-regulate its expression rather than abrogating it.
A novel method has been recently described for selectively reducing the level of target proteins, called Pharmacological Protein Inactivation by Folding Intermediate Targeting (PPI-FIT)\textsuperscript{\tiny{18}}. The technology is based on the concept of targeting folding intermediates of proteins rather than native conformations and is made possible by computational algorithms allowing the all-atom reconstruction of folding pathways\textsuperscript{\tiny{18,19}}. The rationale underlying the PPI-FIT method is that stabilizing a folding intermediate of a protein with small ligands should promote its degradation by the cellular quality control machinery, which could recognize such artificially stabilized intermediates as improperly folded species. Here, we report the full atomistic reconstruction of the folding pathway of ACE2, which predicts the existence of a folding intermediate presenting two unique druggable pockets hidden into the native state. These pockets were employed as novel target sites for virtual screening campaigns aimed at repurposing compounds capable of modulating the expression of ACE2. The ability of virtual hits compounds of reducing the expression of ACE2 was tested in Vero cells. The entry of a pseudotyped retrovirus exposing the coronavirus spike protein was challenged by the presence of the ACE2-suppressing compounds. Most importantly, the antiviral activity of these compounds against live virus has been tested.

\section*{Results}
\subsection*{All-Atom Reconstruction of the ACE2 Folding Pathway}
To obtain the all-atom reconstruction of the folding pathway of apo-ACE2, we employed the so-called Bias Functional approach (see Methods)\textsuperscript{\tiny{20}}. This enhanced path sampling scheme requires to provide the protein native conformation, which was retrieved from PDB 1R42 (Figure 1).

\begin{figure}[!h]
\centering
\includegraphics[scale=0.145]{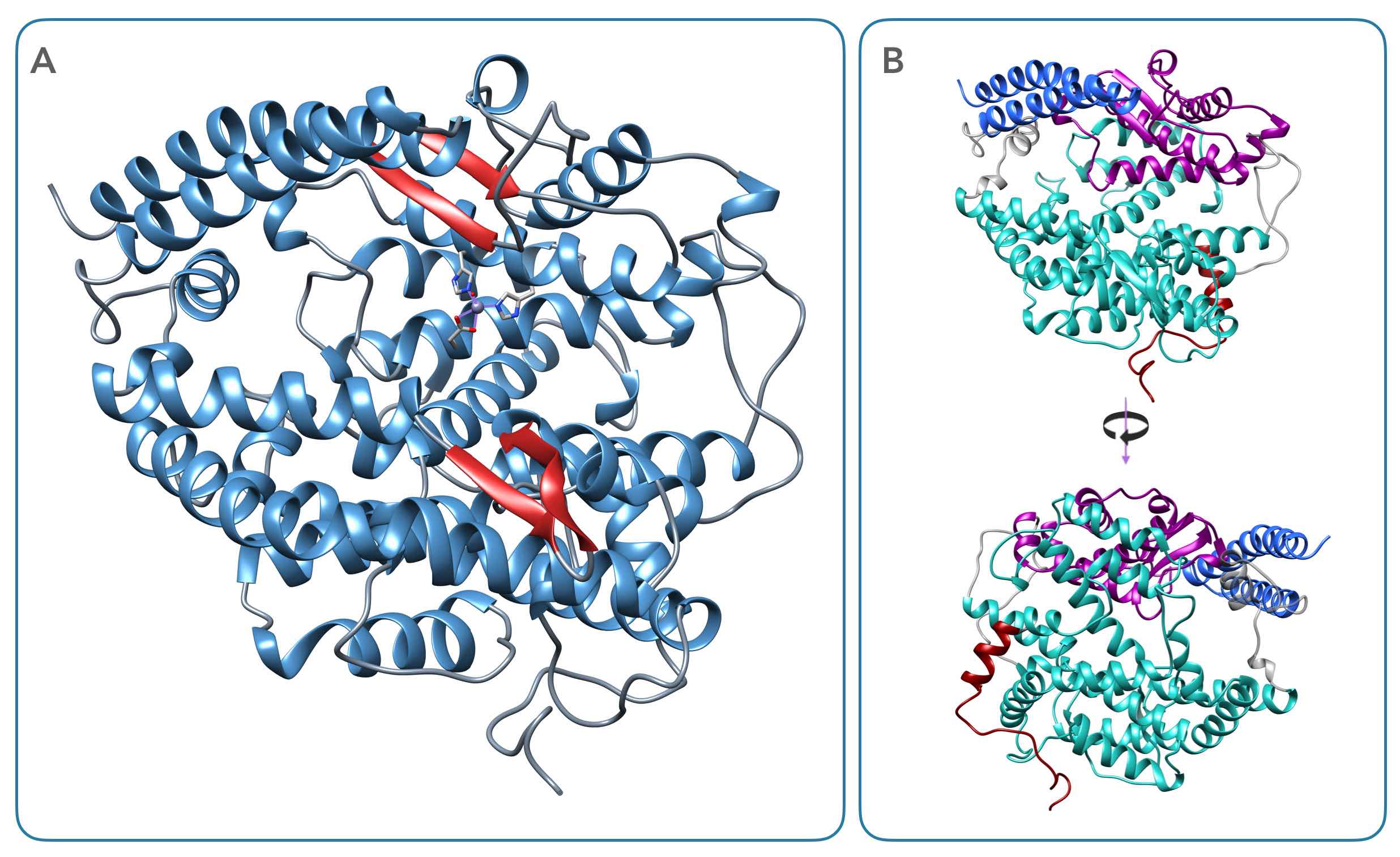}
\caption{\footnotesize{\textbf{Structure of the catalytic domain of ACE2.} A. Cartoon representation of the ACE2 catalytic domain retrieved from PDB 1R42. $\alpha$-helices are depicted in blue, while $\beta$-sheets are represented in red. B. ACE2 cartoon representation from two different perspectives. Folding regions are color-coded as follows: N-terminal region (residues 19-82), blue; central core (residues 149-284 and 430-587), cyan; globular domain coordinating the zinc ion (residues 295-425), magenta; C-terminal tail (residues 589-615), red.}}
\label{Fig1}
\end{figure}

\noindent This structure includes the atomic coordinates of the catalytic domain of the enzyme (residues 19-615). The zinc ion in the active site, as well as the glycans, were both removed, due to uncertainties about their binding kinetics. The Charmm36m forcefield\textsuperscript{\tiny{21}} was employed throughout the simulations. The native structure of ACE2 was employed to generate 32 denatured conformations by a combination of unfolding ratchet-and-pawl Molecular Dynamics (rMD) and high temperature MD simulations (see Methods). For each conformation, 40 trajectories were generated by folding rMD, for a total of 1280 trajectories. All the non-productive trajectories were discarded while folding trajectories of the remaining sets (N = 18) were ranked based on the value of their Bias Functional (Equation 5 in Methods). For each set, we selected the least biased trajectory (LBT), corresponding to the folding pathway with the highest probability to occur in the absence of biasing forces\textsuperscript{\tiny{20}}. To gain information about the folding mechanism, we considered a specific collective variable $R$ defined as a normalized linear combination of the fraction of native contacts (Q) and Root Mean Squared Deviation (RMSD) of atomic positions from the native structure. According to our definition, $R$ = 0 corresponds to the fully denatured state of the protein, while $R$ = 1 to the native configuration. The 18 resulting least biased trajectories were employed to generate the transition path energy function $G(R)$ defined in Methods. This quantity provides a lower bound to reaction limiting free-energy barriers that are overcome along the folding process. Therefore, it can be used to infer the existence of folding intermediates. Since the bias functional approach cannot be used to explore the dynamics within the reactant state, we restricted our analysis to the transition region $R$ > 0.55, where significant tertiary-structure content begins to appear. This reactive section of the pathway contains two non-native metastable states, referred to as "early" and "late" intermediates (Figure 2). 

\begin{figure}[!h]
\centering
\includegraphics[scale=0.29]{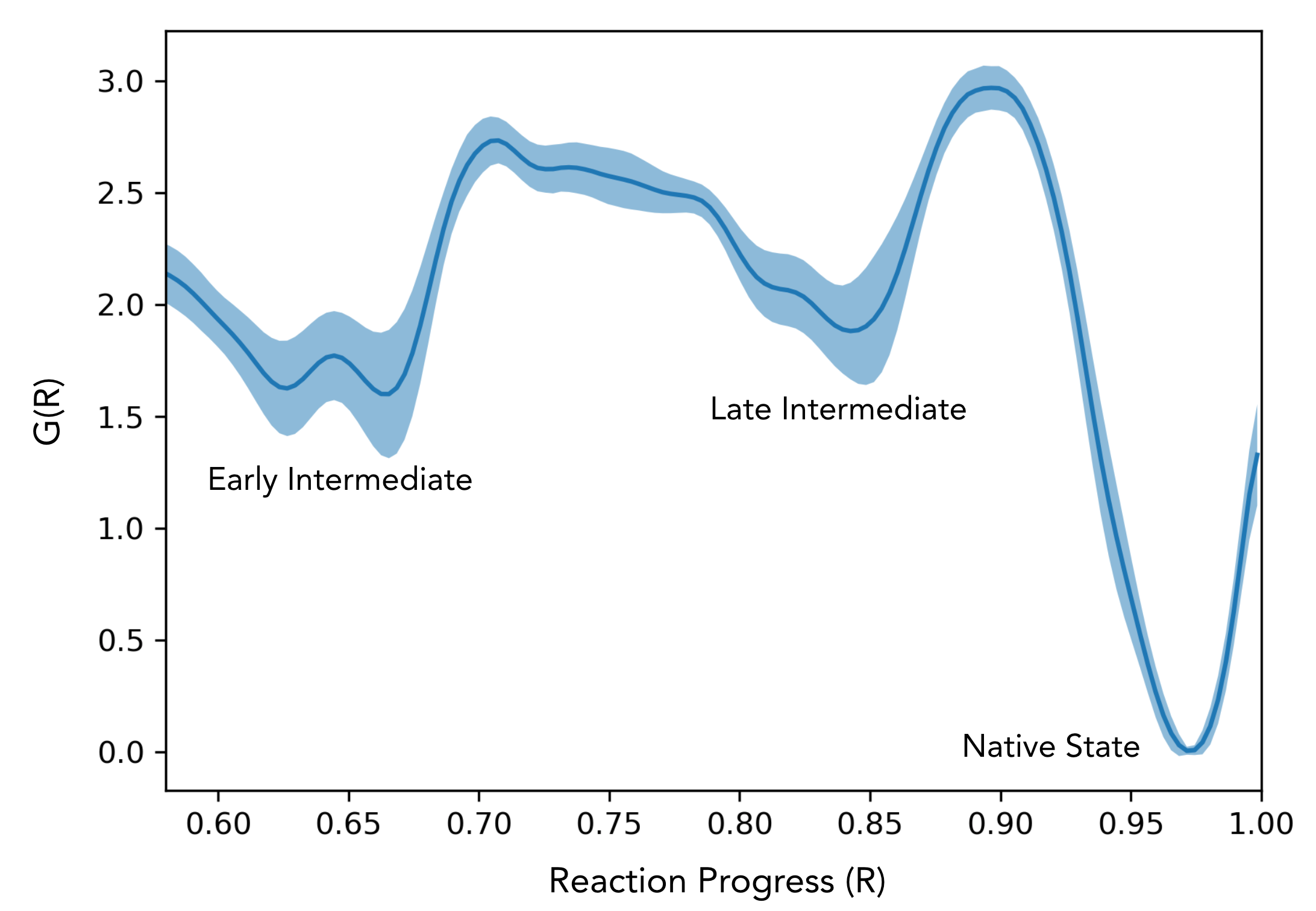}
\caption{\footnotesize{\textbf{Transition path energy profile of the ACE2 folding pathway. }Transition path energy of the ACE2 folding free energy profile, computed as $G(R) = -\text{ln}P(R)$. The reaction coordinate R is a normalized linear combination of the collective variables Q and RMSD. The filled curve represents the standard deviation of the profile, computed using jackknife resampling. The graph shows the presence of two intermediate states (early and late) in addition to the native state. }}
\label{Fig2}
\end{figure}

\noindent The analysis of protein conformations along the trajectories led to the description of the complete folding mechanism for ACE2 (Figure 3). The first event along the ACE2 folding pathway is the formation of three distinct regions (foldons) encompassing: (i) the two N-terminal $\alpha$-helices (residues 19-82); (ii) the globular domain (residues 295-425) containing the zinc coordinating amino acids; (iii) the central core (residues 149-284 and 430-587), followed by the docking of the C-terminal tail (residues 589-615). These regions then sequentially fold onto each other, giving rise to the native state. 

\begin{figure}[!h]
\centering
\includegraphics[scale=0.5]{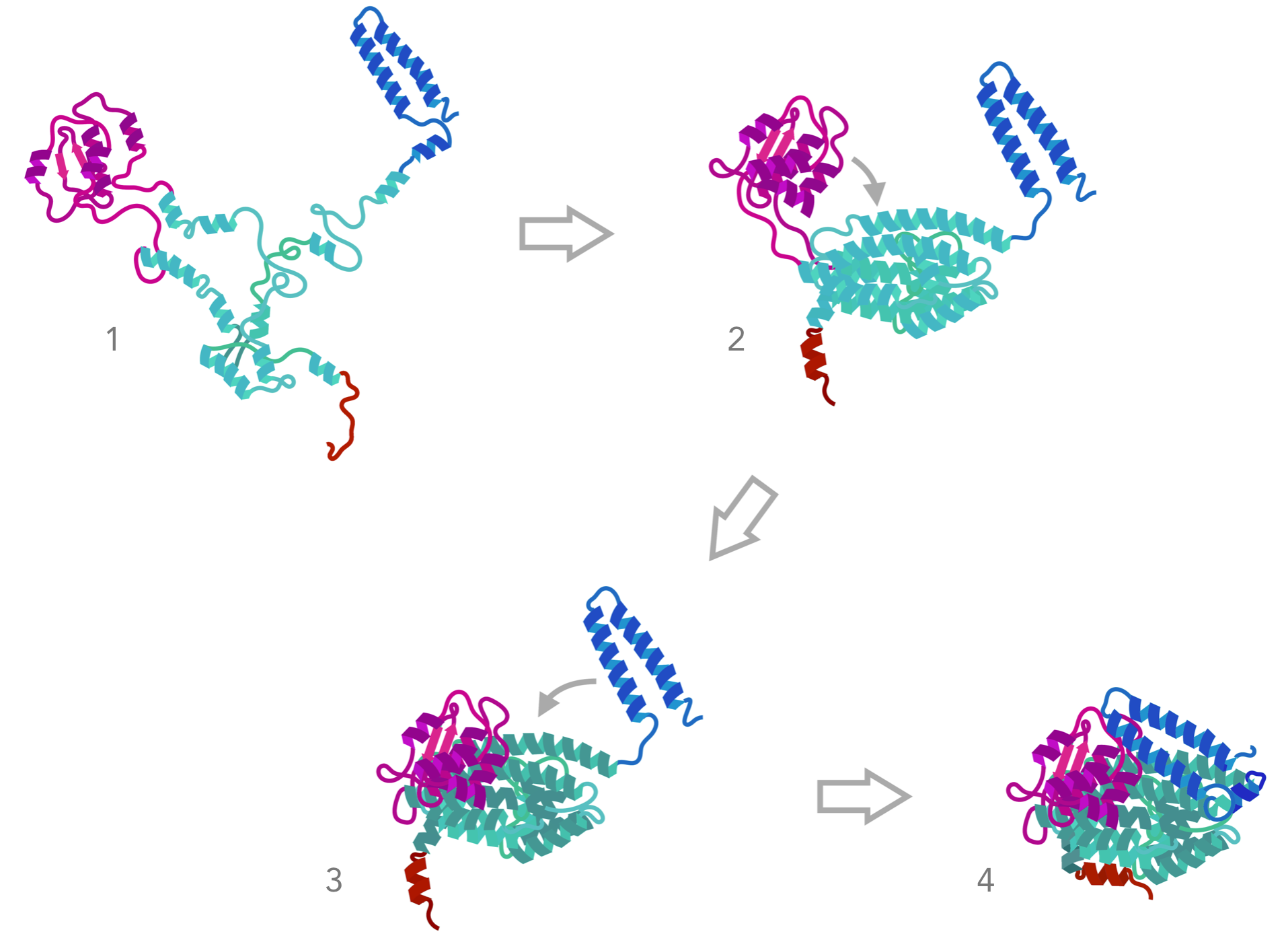}
\caption{\footnotesize{\textbf{Simplified representation of ACE2 folding pathway. } The folding pathway of ACE2 begins with the independent formation of the 3 main foldons (1): the N-terminal region (residues 19-82), depicted in blue; the central core (residues 149-284 and 430-587), represented in cyan; the globular domain coordinating the zinc ion (residues 295-425). Subsequently, the globular domain docks onto the central core (2), followed by the docking of the N-terminal region and the formation of the C-terminal helix (residues 589-598) (3), which then dock on the central core (4), giving rise to the native structure. }}
\label{Fig3}
\end{figure}

\subsection*{Indentification of ACE2 Folding Intermediates}
A k-mean clustering was performed to obtain representative protein structures from the two non-native free energy wells (i.e. early and late intermediates) appearing along the folding pathways (Supp. Figure 1). Four cluster centers per well were selected as representative conformations (Figure 4).  The early intermediate is characterized by the lack of docking of the three foldons, which may also appear as not completely structured in some conformations. Conversely, the late ACE2 folding intermediate differs from the native conformation mainly for the topology of the two N-terminal $\alpha$-helices, which are not docked to the rest of the structure in all the conformations.

\begin{figure}[!h]
\makeatletter
\renewcommand{\fnum@figure}{\small{Figure 4}}
\makeatother
\centering
\includegraphics[scale=0.59]{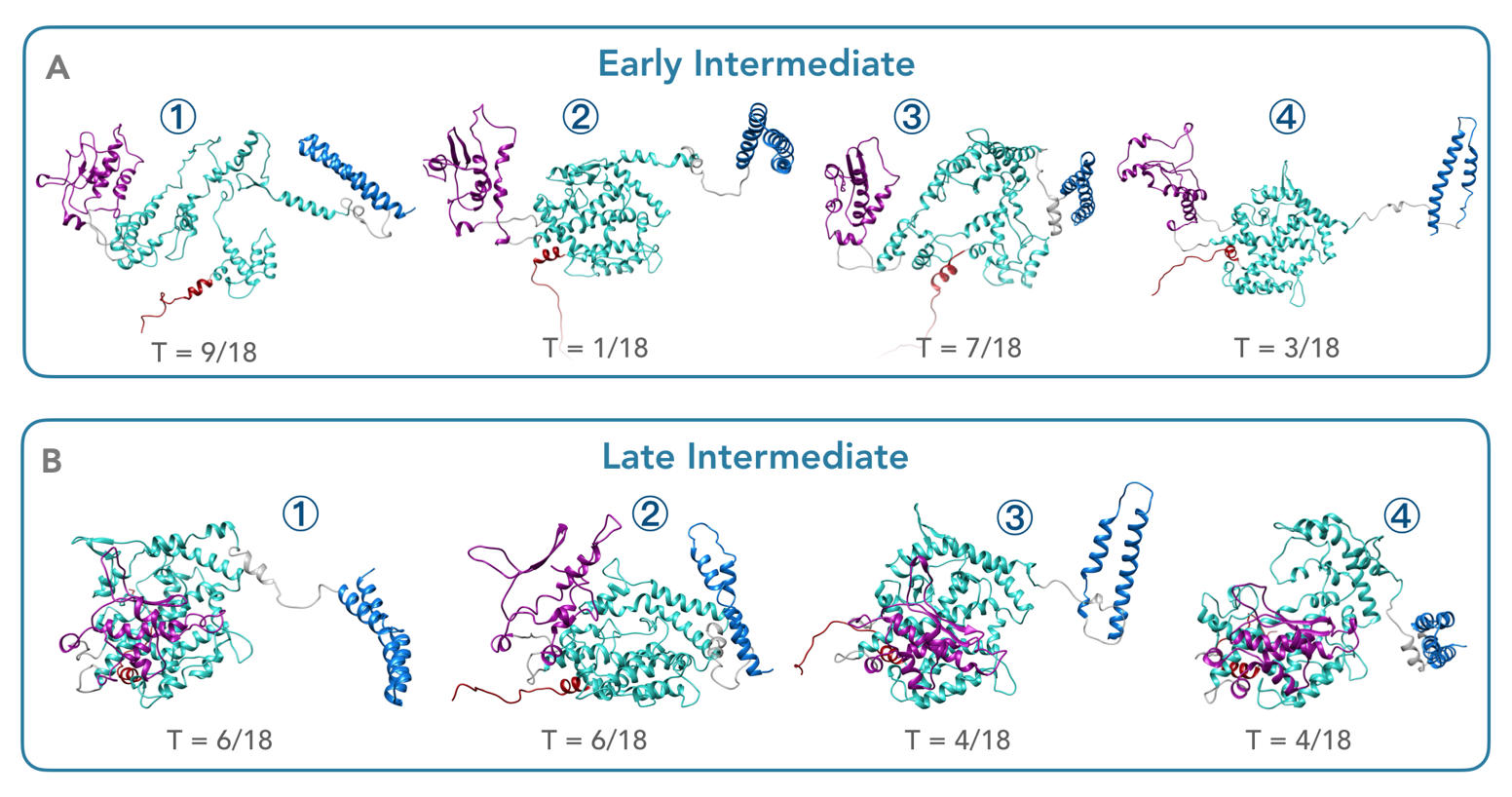}
\caption{\footnotesize{\textbf{Representative folding intermediates conformations. }Representative protein conformations corresponding to the cluster centers of the early (A) and late (B) intermediates. The frequency of the least biased trajectories (T) corresponding to the cluster represented by each protein conformation is shown below each structure. Domains are colored as follows: N-terminal region (residues 19-82), blue; central core (residues 149-284 and 430-587), cyan; globular domain coordinating the zinc ion (residues 295-425), magenta; C-terminal tail (residues 589-615), red. }}
\label{Fig4B}
\end{figure}

\noindent Such a high degree of structural organization and low conformational variability represent ideal features of a folding intermediate, allowing the reliable identification of druggable sites. Thus, the late folding intermediate of ACE2 was further scouted for the presence of pockets suitable for virtual screening campaigns. 

\subsection*{Identification of Druggable Pockets on the ACE2 Late Intermediate.}
\noindent To account for minor variations among the structures belonging to the same cluster, we selected an additional protein conformation for each cluster of the late intermediate (Supp. Figure 2). This process yielded a final number of 8 protein conformations, which were then used for the subsequent pocket detection step. These structures were subjected to 50 ns of MD simulations at T = 310 K with position restrain on C$\alpha$, to explore the conformational variability of the residues side-chains. Then, for each MD simulation, 200 structures, equispaced in time, were extracted and analyzed with the SiteMap\textsuperscript{\tiny{22}} and DogSiteScorer\textsuperscript{\tiny{23}} software. Pockets were then ranked based on a series of druggability descriptors (Supp. Table 1). This analysis identified two predicted druggable pockets, exclusively present in the ACE2 late intermediate state but not in the native conformation (Figure 5).

\vspace{0.2cm}
\begin{figure}[!h]
\makeatletter
\renewcommand{\fnum@figure}{\small{}}
\makeatother
\centering
\includegraphics[scale=0.545]{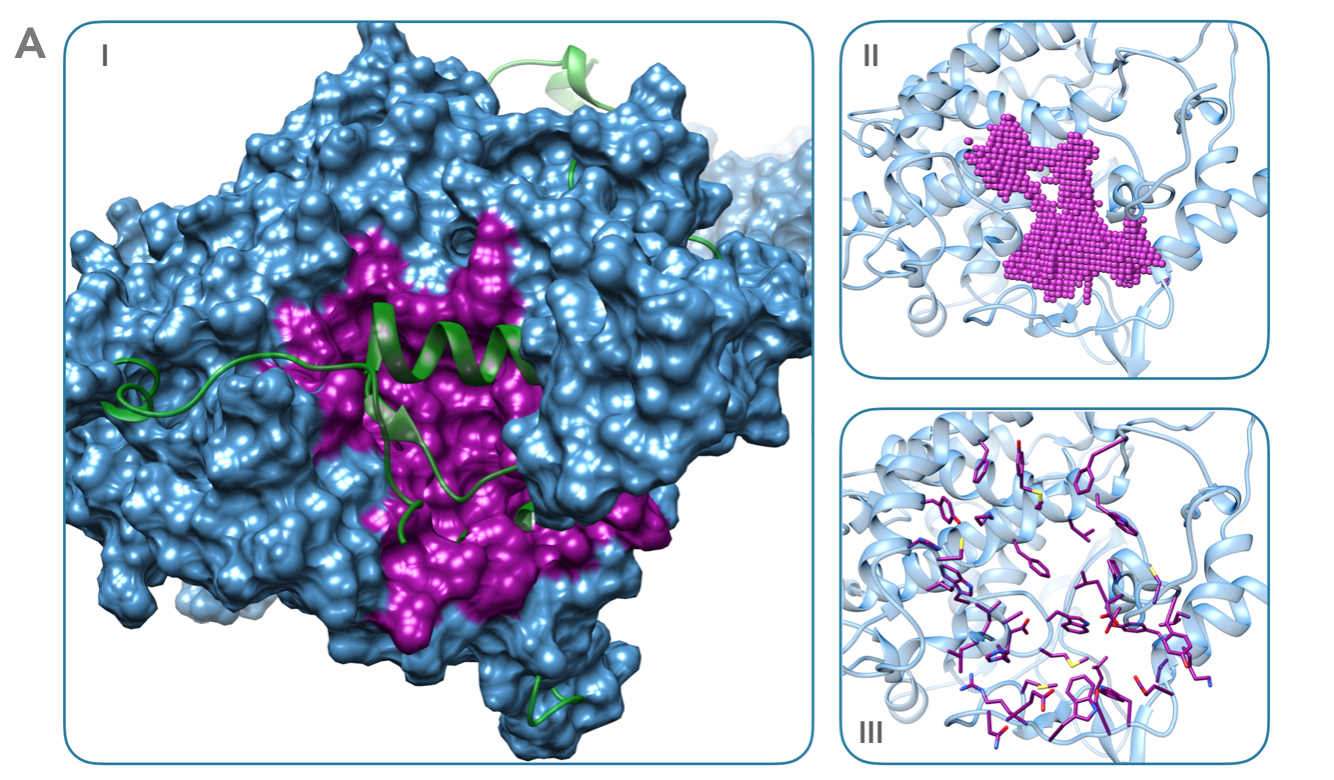}
\caption{}
\label{Fig5A}
\end{figure}

\newpage

\begin{figure}[!h]
\makeatletter
\renewcommand{\fnum@figure}{\small{Figure 5}}
\makeatother
\centering
\includegraphics[scale=0.545]{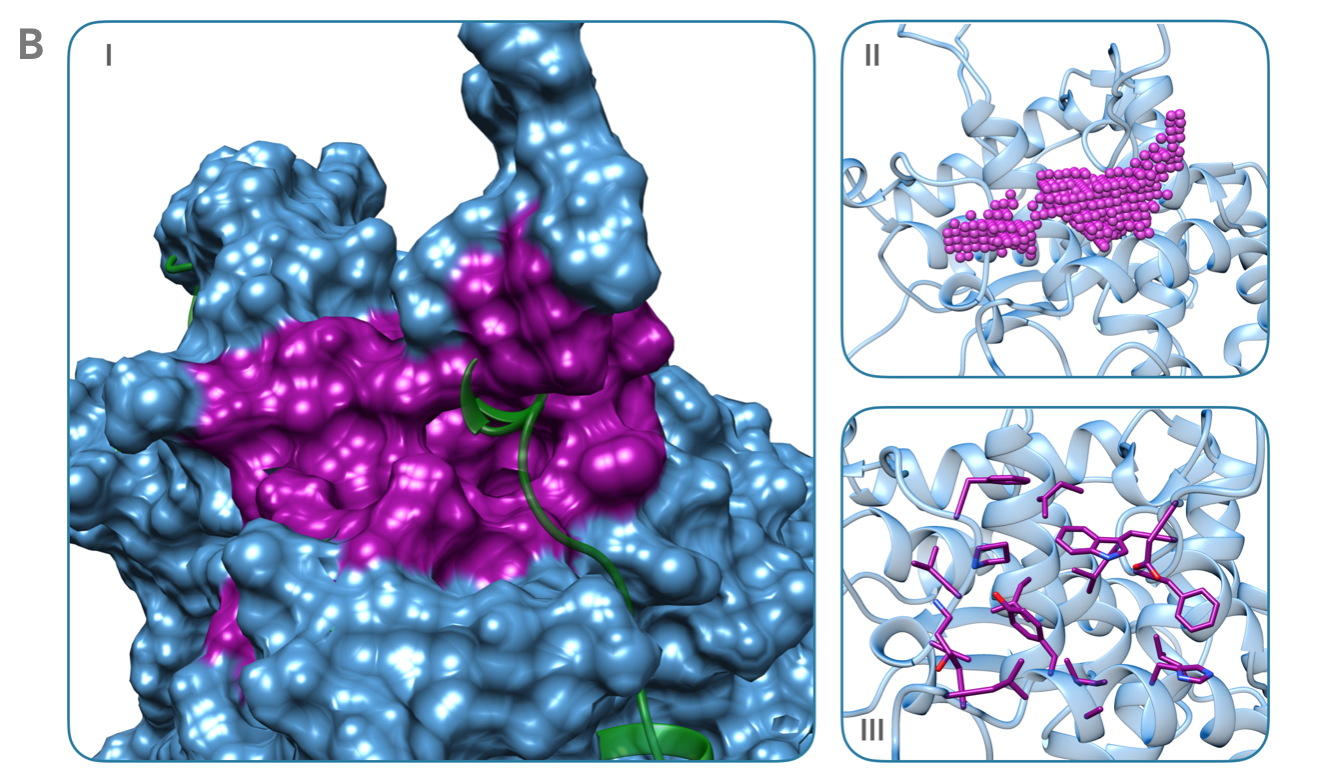}
\caption{\footnotesize{\textbf{Druggable pockets identified in the intermediate state. }A. Druggable pocket 1. Panel I shows the surface of the ACE2 folding intermediate (blue), while the surface of the druggable pocket is shown in magenta. The superimposed structure of the native state is depicted as a green cartoon. In the native state, the structure composed by residues 475-492 covers the binding pocket. In panel II, the intermediate structure is represented as a light blue cartoon, while the volume of the druggable site is colored in magenta. In panel III, the residues forming the identified site are depicted in magenta: 127, 130, 144, 152, 159, 160, 161, 163, 167, 168, 171, 172, 173, 174, 176, 230, 237, 241, 261, 262, 264, 265, 266, 267, 268, 269, 270, 271, 272, 275, 448, 451, 452, 454, 455, 456, 459, 464, 497, 498, 499, 500, 502, 503. B. Druggable Pocket 2. Panel I represents the surface of the intermediate state in blue, while the surface of the druggable pocket is shown in magenta. The superimposed structure of the native state is depicted as a green cartoon. In the native state, the structure composed by residues 598-603 covers the binding pocket. In panel II, the intermediate structure is represented as a light blue cartoon, while the volume of the druggable site is colored in magenta. In panel III, residues forming the identified site are depicted in magenta: 236, 239, 240, 242, 243, 247, 248, 281, 282, 283, 284, 285, 286, 436, 440, 443, 591, 592, 593, 594, 596, 597, 600.  }}
\label{Fig5B}
\end{figure}

\noindent In particular, pocket 1 results from a local structural variation into the core region (residues 468-498), while pocket 2 results from the missing docking of the C-terminal tail onto the core region. To determine the reliability of each druggable site, we measured the number of folding pathways presenting the relevant structural variation underlying each pocket (Figure 6).  

\begin{figure}[!h]
\makeatletter
\renewcommand{\fnum@figure}{\small{Figure 6}}
\makeatother
\centering
\includegraphics[scale=0.17]{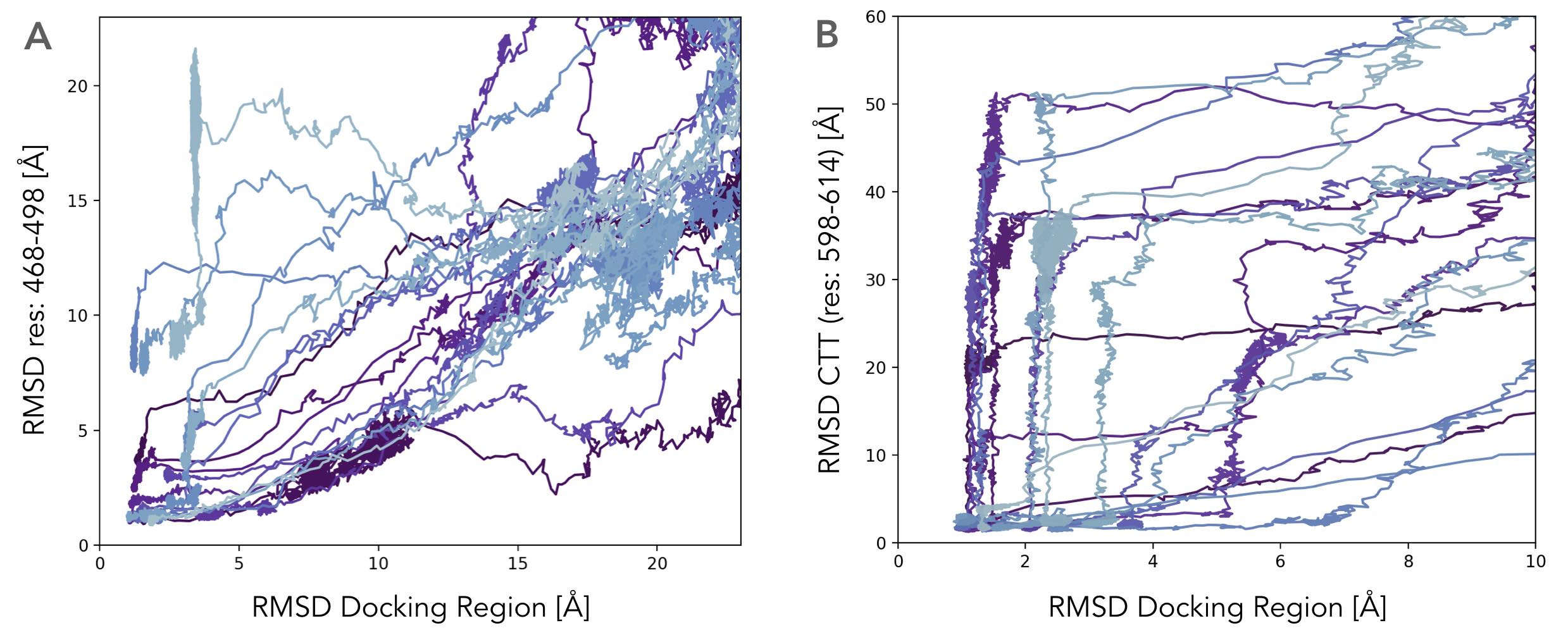}
\caption{\footnotesize{\textbf{Analysis of folding mechanism yielding to pockets formation.} A. The RMSD of the region composed by residues 468-498, which hides pocket 1 in the native structure, is plotted as a function of the RMSD of its docking region on the central core (residues 144-171, 221-283, 442-467, 499-519). In 1/18 trajectories these two regions are distant enough to allow the formation of pocket 1 (RMSD res: 468-498 > 15 Å, RMSD docking region < 4 Å), while in the remaining trajectories they fold cooperatively (15/18) or non-cooperatively (2/18), in both cases not leading to the formation of pocket 1. B. The RMSD of the C-terminal tail (residues 589-615) is plotted as a function of the RMSD of its docking on the central core (residues 221-265, 432-465, 513-533). Pocket 2 appears in all the trajectories (9/18) in which the docking region is structured before the attachment of the C-terminal tail (RMSD CTT > 20 Å, RMSD docking region < 4 Å).}}
\label{Fig6}
\end{figure}

\noindent The least biased trajectories were projected on two graphs plotting the RMSD of each relevant region (residues 468-498 or C-terminal tail) against the RMSD of the corresponding docking site. These analyses revealed that the pocket 1 is present in a single trajectory, while the pocket 2 is predicted to appear in 9 different trajectories. 

\subsection*{In Silico Identification of Potential Binders of ACE2 Intermediate}
The identification of potential ACE2 folding intermediate ligands was pursued by employing a drug repositioning strategy. We built a unique collection of 9187 compounds by combining libraries of drugs approved by the U.S. Food and Drug Administration (FDA) and molecules at different stages of currently ongoing clinical trials (see Material and Methods). The chemical collection was screened against the two identified pockets by following a consensus virtual screening workflow (Figure 7). Two different docking software, Glide\textsuperscript{\tiny{22}} and LeadIT\textsuperscript{\tiny{24}}, were employed in parallel to predict the binding affinity of each compound to the ACE2 folding intermediate pockets. Only compounds showing promising predicted affinity (i.e. Glide\textsubscript{ds} $\leq$ -6 kcal/mol; LeadIT HYDE\textsubscript{aff} $\leq$ 50 $\mu$M) in both docking protocols were submitted to a third docking round based on AutoDock\textsuperscript{\tiny{25}}. This process identified two consensus sets (AD\textsubscript{LBE} $\leq$ -6 kcal/mol, AD\textsubscript{NiC} $\geq$ 25), including 145 compounds for pocket 1 and 238 for pocket 2. The top scoring compounds from Glide (Glide\textsubscript{ds} $\leq$ -9 kcal/mol) and LeadIT (HYDE\textsubscript{aff} $\leq$ 5 $\mu$M) were also added to these sets. Finally, a visual inspection of binding mode and chemical similarity annotation for each ligand allowed the selection of 14 virtual hits for pocket 1 and additional 21 for pocket 2 (Supp. Table 2). 

\begin{figure}[!h]
\makeatletter
\renewcommand{\fnum@figure}{\small{Figure 7}}
\makeatother
\centering
\includegraphics[scale=0.37]{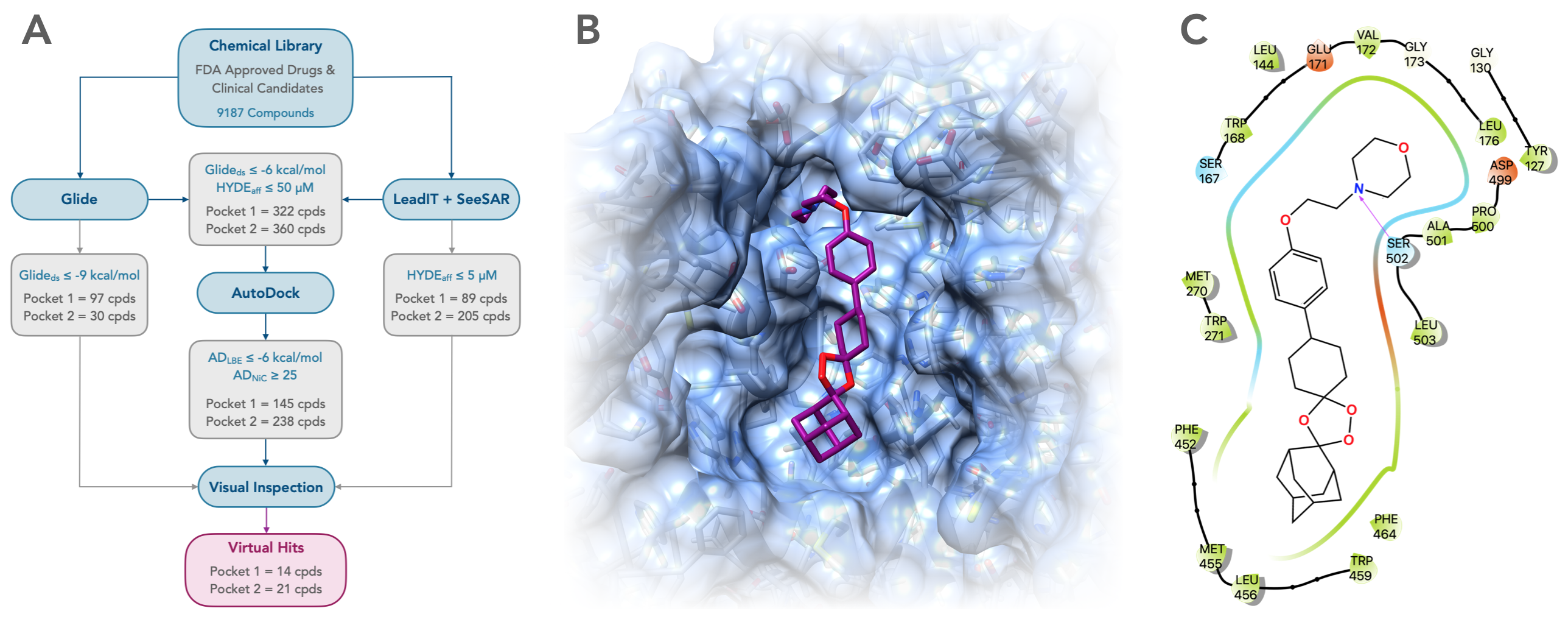}
\caption{\footnotesize{\textbf{Virtual Screening.} A. Schematic of the virtual screening workflow employed for drug repositioning. Three-dimensional binding pose (B) and two-dimensional ligand interaction scheme (C) for the interaction of artefenomel with the pocket 1 of the ACE2 folding intermediate. Purple arrows indicate H-bonds; green lines indicates the $\pi$-stacking. Residues are labeled with different colors, corresponding to negatively charged (red), polar (cyan) and hydrophobic (green).}}
\label{Fig7}
\end{figure}

\noindent Collectively, these results predicted 35 potential ligands for the ACE2 folding intermediate (Supp. Table 2). Out of these 35 predicted ligand, 8 (ALK-4290, Iferanserin, Lifibrol, LY-2624803, PF-00217830, Phenindamine, Serdemetan and Vapitadine) were not commercially available. Instead, we tested 8 additional analogues of mefloquine (Hydroxychloroquine, Piperaquine, Chloroquine, Primaquine, Amodiaquine, Halofantrine, Tafenoquine and Amodiaquine), drug belongs to a class of antimalaria agents recently described for their potential effect against SARS-CoV-2 \textsuperscript{\tiny{26,27}}. While the precise mechanism by which chloroquine and its more active derivative hydroxychloroquine inhibit virus replication is not known, reports suggest that the compounds may act by reducing the glycosylation of ACE2\textsuperscript{\tiny{28}}.


\newpage

\subsection*{Cell-Based Validation of In Silico-Predicted ACE2 Suppressing Drugs}
As a preliminary step, we sought to test the endogenous expression of ACE2 in different untrasfected cells, including HEK293, HepG2, Huh-7 and Vero, by western blotting. Results show that ACE2 expression is higher in Vero cells (Supp. Figure 3). These cells were then employed in subsequent experiments. In order to test the effect of the different molecules on ACE2 expression, Vero cells were exposed for 48 h to each compound at three to five different concentrations (0.3-300 \si{\micro}M, depending on their solubility) and the expression of the protein was evaluated by western blotting. Similarly, cells exposed to the same compounds and concentrations were analyzed by MTT to estimate the intrinsic cytotoxicity of each molecule. Among the in silico predicted 35 different compounds, we validated 9 molecules for their ability to lower at least 25\%  of the expression of ACE2 (Supp. Table 3) by western blotting, as compared to vehicle controls, in at least one experiment. All the molecules were also tested for their intrinsic toxicity, as assayed by MTT (Supp. Table 3).

\subsection*{Cloroquine and Analogues do not Alter the Levels of ACE2}
Chloroquine, a clinically approved drug effective against malaria, has recently attracted widespread interest as potential therapy for COVID-19. The drug has been reported to interfere with terminal glycosylation of ACE2, which may negatively regulate the virus-receptor binding and inhibit the infection. In our screening, we tested Chloroquine diphosphate as well as several related compounds, including Hydroxychloroquine Sulfate, Tafenoquine succinate, Mefloquine, Primaquine diphosphate, Piperaquine phosphate, Halofantrine hydrochloride and Amodiaquine. Results showed a general increase in the ACE2 signal upon treatment with Chloroquine or related compounds (with the exception of Piperaquine), often accompanied by a decrease in ACE2 molecular weight, possibly confirming alterations in the glycosylation of the protein (Supp. Fig. 4). However, based on the absence of ACE2-lowering effects, these molecules were not considered as promising candidates for further analyses.

\subsection*{Dose-Response Validation of ACE2-Reducing Drugs}
The nine different candidate compounds, capable of reducing ACE2 expression, were further analyzed by dose-response experiments using western blotting and MTT assays (Figure 8). Eight to ten concentration points were chosen, depending on solubility and intrinsic toxicity (0.01-300 \si{\micro}M). Five compounds (Bifonazole, Rimonabant, Ebastine, Piperaquine, Brompheniramine) failed to show a dose-dependent effect and/or showed an intrinsic toxicity in the same of range of concentrations at which the ACE2 lowering effects were observed. Conversely, we found 4 drugs capable of dose-dependently lower the expression of ACE2 in absence of significant cytotoxicity. Ziprasidone (an atypical antipsychotic), showed a dose-dependent lowering effect on ACE2 levels, with an estimated inhibitory concentration at 50\% (IC\textsubscript{50}) of 10.6 \si{\micro}M, while the compound showed no significant toxicity even at the highest concentration tested (300 \si{\micro}M). Buclizine (an antihistamine piperazine derivative), significantly reduced ACE2 levels starting from 30 \si{\micro}M, and induced evident cytotoxicity (>30\%) only at 100 \si{\micro}M. Beclabuvir (an antiviral drug for the treatment of hepatitis C virus, currently in clinical trials) was able to dose-dependently reduce ACE2 levels starting from 0.3 \si{\micro}M, and showed a significant toxicity only at 30 \si{\micro}M. Artefenomel (a novel antimalarial trioxolane currently in clinical trials) lowered ACE2 expression in a statistically significant fashion starting from 0.3 \si{\micro}M without inducing any cytotoxicity at the tested concentrations (up to 300 \si{\micro}M). Based on their lowering effects on ACE2, as well as on the evaluation of their intrinsic toxicity, we elected these compounds as positive hits for further analyses. 

\subsection*{Evaluation of Inhibitory Effects on a Pseudotyped Retrovirus Exposing the SARS-CoV-2 Spike Protein}
The SARS-CoV-2 virus gains entry into various cells through the binding of its surface protein spike to the host ACE2 receptor. Thus, compounds capable of lowering the expression of ACE2 should also be able to prevent virus entry into cells. In order to directly test this possibility, we employed a pseudotyped retroviral vector exposing the SARS-CoV-2 spike protein, as well as a GFP reporter.

\begin{figure}[!h]
\makeatletter
\renewcommand{\fnum@figure}{\small{Figure 8}}
\makeatother
\centering
\includegraphics[scale=0.475]{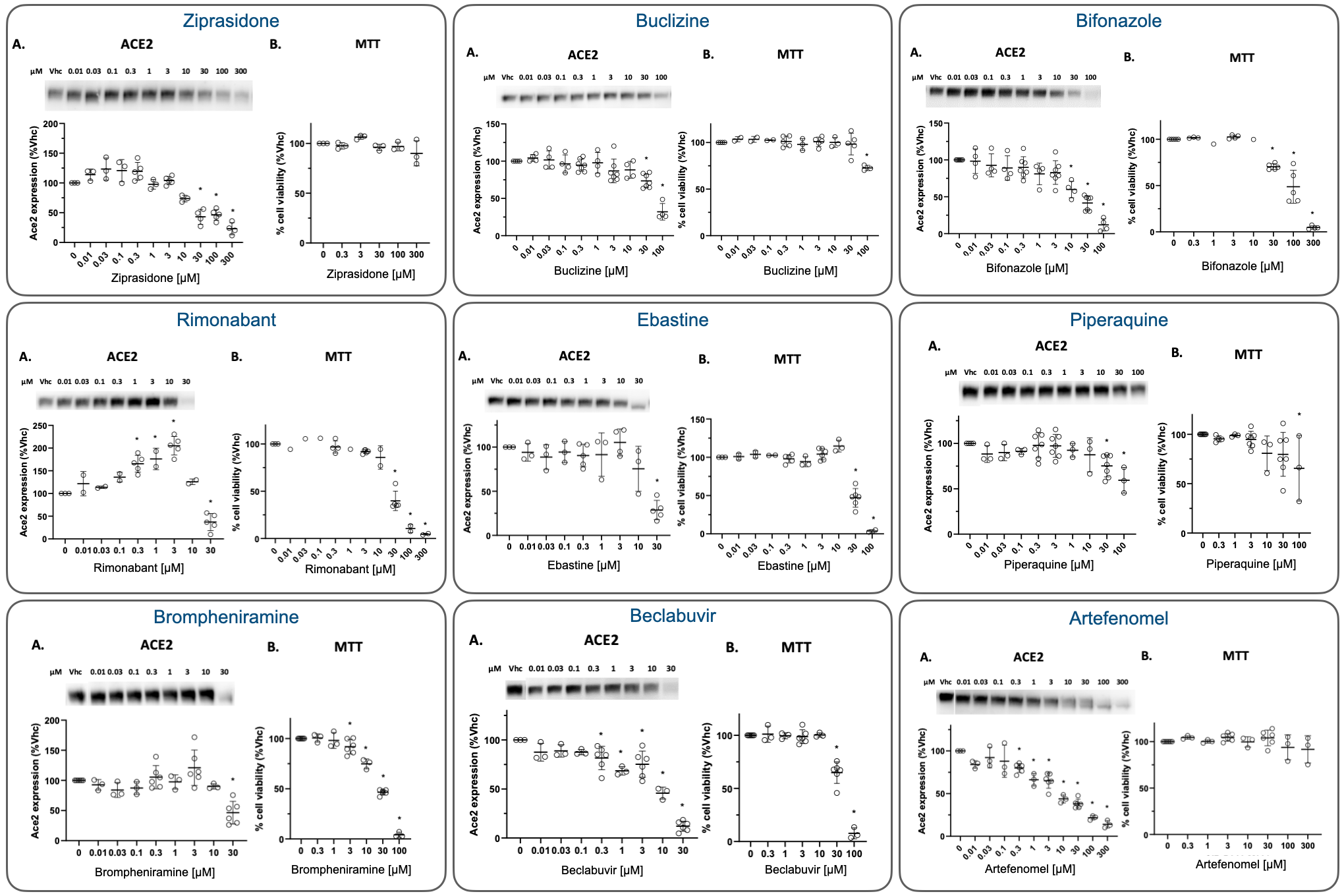}
\caption{\footnotesize{\textbf{Cell-based validation of candidate hits.} Untransfected Vero cells were exposed to different concentrations of each compound (indicated) or vehicle (DMSO or Milli-Q or Methanol, volume equivalent) for 48 h, lysed and analyzed by western blotting. Signals were detected by using specific anti-ACE2 primary antibody, relevant HRP-coupled secondary antibodies, and revealed using a ChemiDoc Touch Imaging System. Western blot images are representative examples of different experiments (n $\geq$ 3). The graphs show the densitometric quantification of the levels of ACE2 (A). Each signal was normalized on the corresponding total protein lane (detected by UV, and allowed by the enhanced tryptophan fluorescence technology of stain-free gels) and expressed as the percentage of the level in vehicle (Vhc)-treated controls. B. The intrinsic toxicity of each molecule was assessed by MTT assay. The graphs show cell viability values expressed as percentage of vehicle (DMSO or MilliQ-water, volume equivalent)-treated cells. Concentration points were chosen depending on solubility and intrinsic toxicity. None of the compounds show toxicity at the indicated concentration. Statistically significant differences are indicated by the asterisk (* p < 0.05). }}
\label{Fig8}
\end{figure}

\vspace{0.5cm}

\noindent Vero cells incubated with each of the four candidate compounds at different concentrations were transduced with retroviral vectors pseudotyped with the SARS-CoV-2 spike protein, or with control vectors without it. The effect of each compound on retroviral vector transduction was estimated by quantifying the relative percentage of cells presenting the GFP fluorescence. We found that all the four compounds inhibited retroviral transduction in a dose-dependent fashion, at concentrations similar to those at which the molecules lowered ACE2 expression (Figure 9). Importantly, none of the compounds induced significant cytotoxicity in this assay, with the exception of Beclabuvir, which showed cytotoxicity but only at the highest concentrations tested (30 and 100 \si{\micro}M, not shown). Collectively, these results indicate that the ability of the selected compounds to lower the expression of ACE2 translates in a reduced cellular entry for a pseudotyped retroviral vector exposing the SARS-CoV-2 spike protein.

\newpage

\begin{figure}[!h]
\makeatletter
\renewcommand{\fnum@figure}{\small{Figure 9}}
\makeatother
\centering
\includegraphics[scale=0.53]{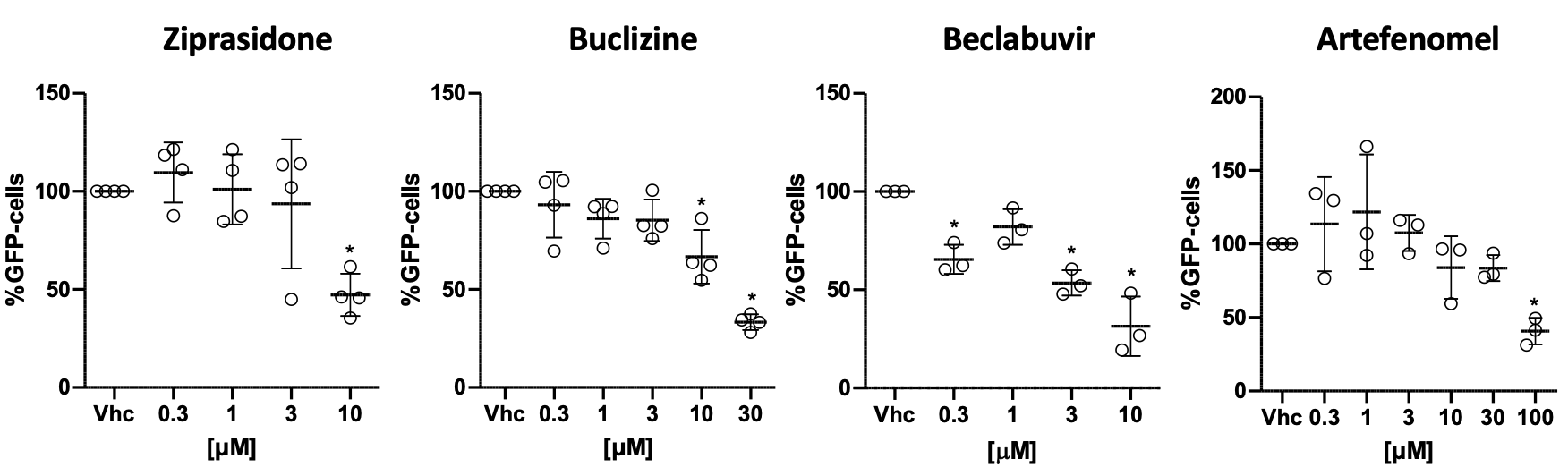}
\caption{\footnotesize{\textbf{Effect of ACE2-lowering drugs on the transduction efficiency of a pseudotyped retroviral vector.} Vero cells exposed to each compound at the indicated concentration were transduced with a  SARS-CoV-2-Spike protein pseudotyped retroviral vector   functionalized with a GFP reporter gene. Identical retroviral vectors missing the spike protein were used as controls. The number of transduced cells were quantified by detecting the GFP fluorescence using a plate reader and analyzed with the ImageJ software (NIH). The number of fluorescent cells was normalized to the amount of cells within each well, estimated by using the MTT assay, and expressed as the percentage of the vehicle control. For each condition, mean $\pm$ SD were calculated from at least 3 independent replicates. Statistical analyses were performed using the one-way ANOVA Dunnett's post-hoc test. Each compound was tested at relevant concentrations excluding those at which the molecule showed detectable intrinsic fluorescence. Significant changes are indicated by an asterisk (* p < 0.05).}}
\label{Fig9}
\end{figure}

\subsection*{Antiviral activity against live SARS-CoV-2}
Sibylla Biotech SRL requested RetroVirox Inc., San Diego, California, to perform full-dose antiviral testing on the four candidate compounds. Assays against live SARS-CoV-2 were performed against the MEX-BC2/2020 strain. A cytopathic effect (CPE) based antiviral assay was performed by infecting Vero E6 cells in the presence or absence of test-items. Infection of cells leads to significant cytopathic effect and cell death after 4 days of infection. In this assay, reduction of CPE in the presence of inhibitors was used as a marker to determine the antiviral activity of the tested items (Figure 10). 

\begin{figure}[!h]
\makeatletter
\renewcommand{\fnum@figure}{\small{Figure 10}}
\makeatother
\centering
\includegraphics[scale=0.362]{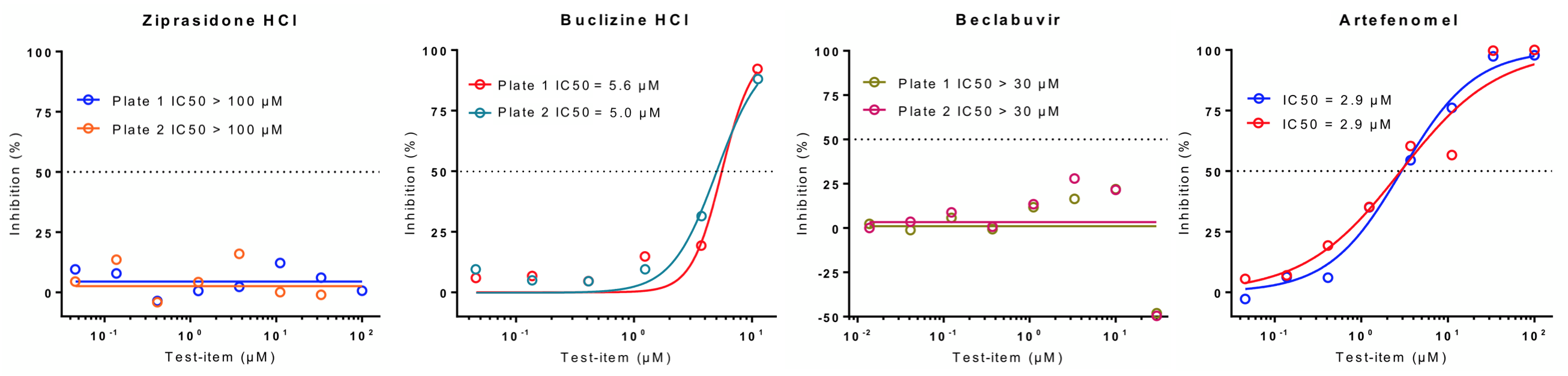}
\caption{\footnotesize{\textbf{IC\textsubscript{50} values for Inhibition of SARS-CoV-2 CPE by test-items.} Values indicate the percentage inhibition of the CPE induced by live SARS-CoV-2 (MEX-BC2/2020), as compared to samples incubated with no test-item (vehicle alone). Results show the average of duplicate data points from two separate plates for Ziprasidone, Buclizine and Beclabuvir, and the average of triplicate data points from two separate plates for Artefenomel. Data was modeled to a sigmoidal function using GraphPad Prism software fitting a normalized dose-response curve with a variable slope.}}
\label{Fig10}
\end{figure}

\noindent Viability assays to determine test-item-induced loss of cell viability was monitored in parallel using the same readout, but treating uninfected cells with the test-items. Antiviral and cytotoxic effects (expressed as IC\textsubscript{50} and CC\textsubscript{50}) are summarized in Table 1. Of the four test-items evaluated, Artefenomel completely prevented the virus-induced CPE in the concentration range 33 \si{\micro}M to 100 \si{\micro}M, resulting in viability levels similar to those observed in uninfected cells. The antiviral activity of Artefenomel shows a dose-response curve with an IC\textsubscript{50} of 2.9 \si{\micro}M. The cell viability assay further assessed that the antiviral activity displayed by Artefenomel was not due to cytotoxicicity. None of the concentrations evaluated of the test-item displayed any cytotoxicity. Buclizine also completely prevented the virus-induced CPE at a concentration of 11 \si{\micro}M, where the cell viability resulted in 75\% of vehicle. The dynamic range of the antiviral activity displayed by Buclizine was narrow, with an IC\textsubscript{50} value between 5 and 6 \si{\micro}M. Significant inhibition of CPE was only observed between 1 \si{\micro}M and 11 \si{\micro}M, and then disappeared at higher concentrations (data not shown), likely due to the intrinsic cytotoxicity induced at such concentrations, as shown in the absence of SARS-CoV-2. A modest antiviral activity was also observed with Beclabuvir at 1 \si{\micro}M and 10 \si{\micro}M. At a concentration of 30 \si{\micro}M the compound induced cytotoxicity and lost antiviral activity. In the tested condition, Ziprasidone showed no antiviral activity in this assay.

\begin{figure}[!h]
\makeatletter
\renewcommand{\fnum@figure}{\small{Table 1}}
\makeatother
\centering
\includegraphics[scale=0.47]{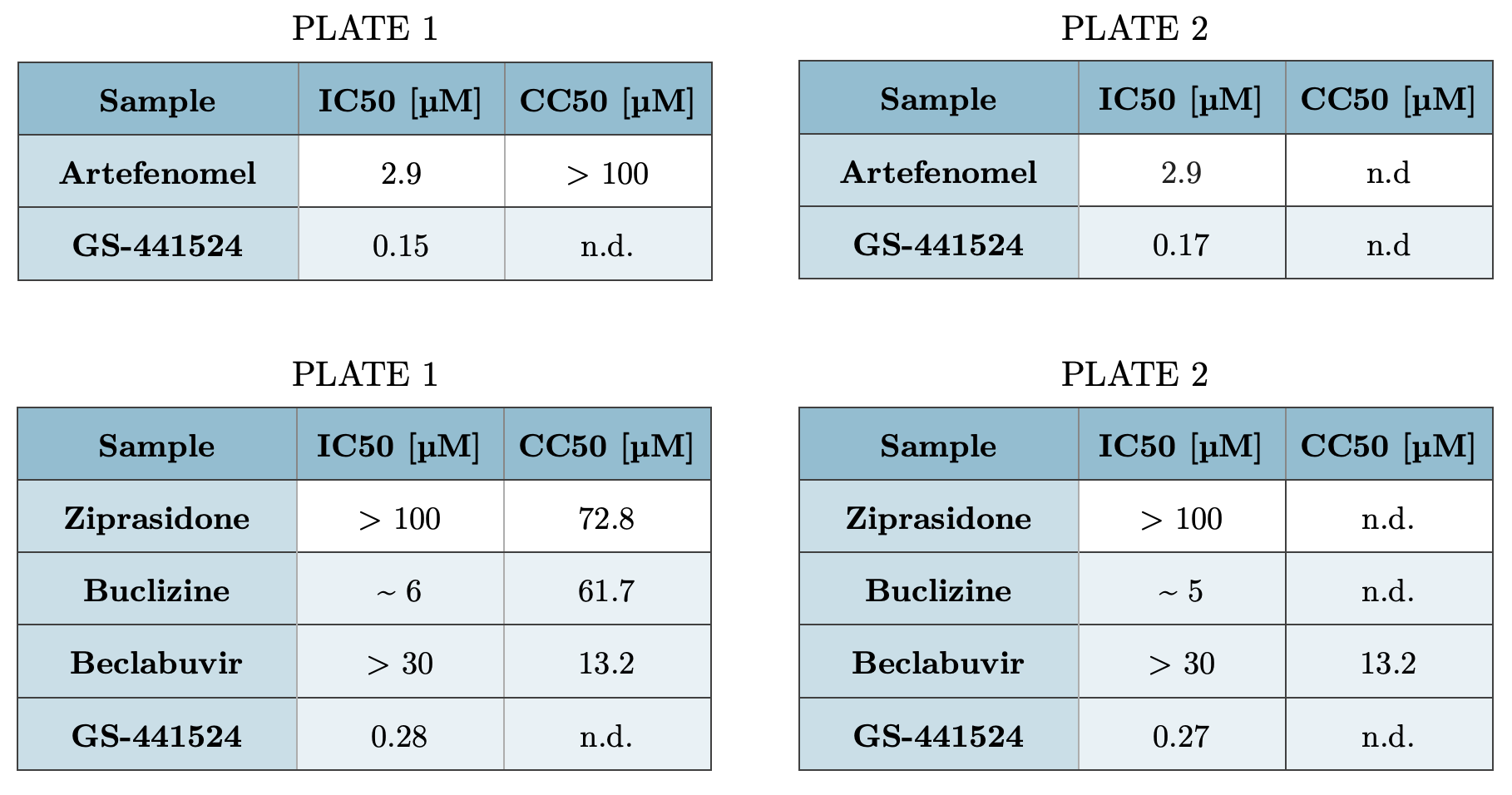}
\caption{\footnotesize{\textbf{Summary of antiviral and cytotoxic results.} IC\textsubscript{50} (antiviral), and CC\textsubscript{50} (cytotoxicity) values are shown for the test-items and GS-441254. Signal-to-background ratios (S/B), and average coefficients of variation (C.V.) are shown. The average CV was determined for all data points in the case of Artefenomel and for duplicate data-points for which 50\% or greater A540 values were observed, as compared to cells infected in the presence of vehicle alone (CPE assay), or uninfected cells (viability assay) in the case of Ziprasidone, Buclizine and Beclabuvir. When viral inhibition, or cell viability (CC50) did not reach 50\% at the highest concentration tested, the IC\textsubscript{50} or CC\textsubscript{50} values are shown as greater than the highest concentration tested. IC\textsubscript{50} values were generated with Graphpad Prism in the case of Artefenomel. IC\textsubscript{50} values for Buclizine could not be calculated with Graphpad Prism software due to the loss of antiviral activity observed at the highest concentrations. The IC\textsubscript{50} value shown was manually extrapolated after eliminating the data-points for 33 \si{\micro}M and 100 \si{\micro}M concentrations.}}
\label{Table1}
\end{figure}

\section*{Discussion}
Multiple pieces of evidence indicate that downregulating the expression of ACE2 of the SARS-CoV2 infection should effectively inhibit virus replication. However, selectively decreasing the expression of a host target protein could be a difficult task. RNA silencing or CRISPR-based strategies represent valid options, but their use could be limited by delivery issues\textsuperscript{\tiny{29,30}}. These problems could be overcome by emerging pharmacological technologies like the proteolysis targeting chimeras (PROTACs), which build on the principle of designing bi-functional compounds capable of interacting with the target protein with one side and engaging the E3 ubiquitin ligase with the other, leading to the degradation of the polypeptide by the proteasome\textsuperscript{\tiny{31}}. Similarly, the PPI-FIT method capitalizes on the cellular quality control machinery to promote the degradation of the target polypeptide, although it does not require the development of bi-functional molecules. PPI-FIT-derived compounds aim at stimulating the removal of the target protein by directly blocking its folding pathway\textsuperscript{\tiny{18}}. In this manuscript, we described the application of the PPI-FIT paradigm to ACE2. Our analyses predict the existence of a folding intermediate showing two unique druggable pockets not present in the native ACE2 conformation. In order to respond to the urgent need for an effective therapy against SARS-CoV-2, we targeted both pockets by an in silico virtual screening approach aimed at repurposing drugs currently in clinical trials or already approved by the FDA. Nine molecules were found to decrease ACE2 expression in Vero cells. Five of those were discarded as showing toxicity in the same concentration range at which ACE2 expression is lowered. The remaining four compounds were effective in reducing ACE2 levels at concentrations showing limited cellular toxicity, if any. The four compounds were able to interfere with the cellular entry of a pseudotyped retrovirus exposing the SARS-CoV-2 spike protein with dose- response profiles closely resembling those observed for their ability in lowering ACE2 expression. The reduction of ACE2 at the cellular level, results in reduced transduction of a pseudotyped virus. Among the four positive molecules, Ziprasidone and Artefenomel showed the highest activity in the absence of any toxicity. Most importantly, the antiviral activity of the four compounds against live SARS-CoV-2 has been tested. Artefenomel has shown inhibition of the cytopathic effect induced by the presence of the live virus with dose-response profile showing an IC\textsubscript{50} of 2.9 \si{\micro}M, with no cytotoxic effect at all concentrations tested. These results strongly support the rationale behind the application of the PPI-FIT methodology to tackle SARS-CoV-2 infection.

\newpage

\section*{Methods}
\subsection*{Software \& Resources}
The simulations were performed on the high-performance-computing facilities of the Italian Institute for Nuclear Physics (INFN) and on the computer cluster of Sibylla Biotech. A modified version of Gromacs 2019\textsuperscript{\tiny{32}} was employed to run the calculations. Data analysis was carried out using Python and its libraries: NumPy, SciPy, Matplotlib and MDAnalysis. 

\subsection*{Structure \& Topology}
The structure of ACE2 was retrieved from PDB 1R42. The file contains the catalytic domain (residues 19-615) solved by X-ray crystallography at 2.2 Å of resolution. Protein topology was generated using Charmm36m after removal of the catalytic zinc ion and sugar moieties. Cysteines were treated in the reduced state. Water molecules were modeled using the Charmm-modified TIP3P. 

\subsection*{Generation and Selection of Unfolded Initial Conditions}
The ACE2 structure was positioned in a cubic box with minimum wall distance equal to 60 Å. The system was filled with water molecules, neutralized with 28 Na$^+$ ions and brought to a final 150 mM NaCl concentration. This setup was followed by an energy minimization using the steepest descent algorithm. An NVT equilibration at 800 K was carried out using the V-rescale thermostat. Then, so-called ratchet and pawl (rMD) molecular dynamics simulations\textsuperscript{\tiny{33,34}} were employed to obtain the denatured states. In this method, an external biasing force is added, acting along a single collective variable, denoted as $z(X)$: 

\begin{equation}
z(X) = \sum_{|i-j|>35}^N \big[  C_{ij}(X) -C_{ij}(X_R)   \big]^2
\end{equation}

\noindent where $C_{ij}(X)$ is an entry of the instantaneous contact map and $C_{ij}(X_R)$ is an entry of the reference contact map, both defined as: 

\begin{equation}
C_{ij}(X) = \frac{1 - (|\textbf{x}_i -\textbf{x}_j |/r_0)^6}{1 - (|\textbf{x}_i -\textbf{x}_j |/r_0)^{10}}
\end{equation}

\noindent where $\textbf{x}_i$ and $\textbf{x}_j$ are the coordinates of the $i_{th}$ and $j_{th}$ atoms of the protein; while $r_0$ is the reference contact-distance that is set to 7.5 \AA. In applications of rMD to protein denaturation, all the entries of the reference contact map were set to 0. In applications of rMD to protein folding simulations, the reference contact map $C_{ij}(X_R)$ is calculated from Eq. (2) using the protein’s native conformation, obtained by performing energy minimization starting from the PDB structure. In any rMD simulation, the system evolves according to plain MD as long as the value of $z(x)$ spontaneously decreases during time. Instead, system fluctuations leading to an increase in $z(X)$ result in the activation of the biasing force defined as:

\begin{equation} 
\textbf{F}_i(X, z_m) = \begin{cases}
-k_R \nabla _i z(X) [z(X) - z_m(t)]  &\text{if $z(X) > z_m(t)$}\\
0 &\text{if $z(X) \leq z_m(t)$}
\end{cases}
\end{equation}

\noindent where $z_m(t)$ is the minimum value assumed by $z(X)$ up to time $t$, and the index $i$ indicates the atom on which the force is acting. To obtain the denatured protein conformations, 32 rMD trajectories of 1 ns each with a force constant $k_R$ = 8 $\cdot$ 10$^{-4}$ kJ/mol. These simulations were followed by 4 ns of plain MD at 800 K in the NVT ensemble. For each plain MD unfolding trajectory, the protein conformation minimizing the following function was selected: 

\begin{equation}
S(X) = \frac{\sum_{i=0}^{n-m} \sum^n_{j=i+m}(j-i)^2\text{max}(1-r_0^{-1}  \Vert \textbf{x}_i - \textbf{x}_j \Vert, 0)}{\sum_{i=0}^{n-m} \sum^n_{j=i+m}(j-i)^2}
\end{equation}

\noindent where $n$ is the total number of atoms, $m$ is the number of ignored subsequent atoms and $r_0$ is a parameter, set to 40 \AA. This metrics is a norm of the contact maps, where atoms with distant indexes weight more in the calculation. Conformations with low value of $S(X)$ are characterized by high degree of denaturation, vice-versa, conformations with high value of $S(X)$ are less denatured.

\subsection*{Generation of the Folding Trajectories}
Selected initial conditions (N = 32) were positioned in a cubic box with minimum wall-distance of 10 \AA. Each system was filled with water molecules, neutralized with 28 Na$^+$ ions and brought to a final 150 mM NaCl concentration. Energy minimization was subsequently performed using the steepest descent algorithm. Then, each system was subjected to 500 ps of NVT equilibration at 350 K (using the V-rescale thermostat\textsuperscript{\tiny{35}}), followed by 500 ps of NPT equilibration at 350 K and 1 Bar (using the V-rescale thermostat and the Parrinello-Rahman barostat). During equilibration, position restraints with force constant 1000 kJ$\cdot$mol$^{-1}$nm$^{-2}$ where introduced on heavy atoms. For each initial condition, 40 rMD trajectories were generated, each one consisting in 5 $\cdot$ 10$^6$ steps with integration time-step of 2 fs. In this set of rMD simulations, reference contact map entering Eq. 1 was calculated using the native structure of ACE2. Cutoff for Coulomb interactions was set to 12 \AA, cutoff for Van der Waals interactions was set to 12 \AA \hspace{0.5mm} with force-switch having 10 \AA \hspace{0.5mm} radius. Long range electrostatics was treated with particle mesh Ewald. Trajectories reaching a configuration with an RMSD lower than 4 \AA \hspace{0.5mm} were considered productive folding events. For each set of folding trajectories (starting from the same initial condition) the pathway with the highest probability to occur in absence of external bias was identified by selecting the one minimizing the Bias Functional  [20], defined as:

\begin{equation}
T = \sum_{i=1}^N \frac{1}{m_i \gamma_i} \int_0^t d \tau \vert \textbf{F}_i^{rMD}(X, \tau ) \vert ^2
\end{equation}

\noindent Where $m_i$ and $\gamma _i$ are the mass and the friction coefficient of the $i_{th}$ atom, $F_i^{rMD}$ is the force acting on it, while $t$ is the trajectory folding time. 

\subsection*{Reconstruction of the Transition Path Energy Profile}
The collective variables Q and RMSD were linearly combined and normalized to obtain a one-dimensional reaction coordinate, $R$:
\begin{equation}
R = \frac{1}{1.34} \Big[ 0.89 \Big( 1 - \frac{RMSD - RMSD_{\text{min}}}{RMSD_{\text{max}}} \Big)  + 0.45 \Big(   \frac{Q - Q_{\text{min}}}{Q_{\text{max}} - Q_{\text{min}}}      \Big)  \Big]
\end{equation}
 
\noindent From the frequency histogram of the least biased trajectories we estimated the probability $P(R)$ to observe a given value of the collective variable. We refer to the quantity $G(R)= - \text{ln}[P(R)]$ as to the “transition path energy”. In a biased dynamics, $G(R)$ yields a lower-bound to the rate limiting free-energy barriers, thus provides a useful tool to identify metastable states. The standard deviation $\sigma$ was estimated through a jackknifing procedure, in particular, the free energy profile was computed for 18 times by leaving a different trajectory out in each calculation. For each bin, the standard deviation $\sigma$ was calculated as:

\begin{equation}
\sigma = \sqrt{\frac{n-1}{n}\sum_{i=1}^n \big( \hat{\theta}_i - \hat{\theta}_{(\cdot)}   \big)}
\end{equation}

\noindent where $n$ is the total number of leas biased trajectories, $\hat{\theta}_i$ is the mean value of the jackknife replicate (where the ith trajectory was removed) and $\hat{\theta}_{(\cdot)}$ is the mean of the jackknife replicates. 

\subsection*{Clustering Analysis}
Protein conformations were sampled from the two identified energy wells, in particular, the early-intermediate was defined with 0.59 < $R$ < 0.68, while the late-intermediate was defined with 0.797 < $R$ < 0.865 (Supp. Figure 1). The two sets of structures were then subjected to k-mean clustering ($k$ = 4), using the Frobenius norm of the contact maps as a distance metrics. Then, for each cluster, a single representative conformation was selected by choosing the structure minimizing its distance from the relative cluster center. 

\subsection*{Identification of the Binding Pockets}
The 4 cluster centers of the late intermediate and 4 additional conformations (each one extracted as a cluster variant) were positioned in a cubic box with minimum wall-distance of 10 \AA, that was filled with water molecules. The system was neutralized with 28 Na$^+$ ions and brought to a final 150 mM NaCl concentration. Energy minimization was subsequently performed, using the steepest descent algorithm. Then, each system was equilibrated for 500 ps in the NVT ensemble at 310 K (using the V-rescale thermostat), followed by 500 ps of NPT equilibration at 310 K and 1 Bar (using the V-rescale thermostat and the Parrinello-Rahman barostat). During equilibration, position restraints with force constant 1000 kJ$\cdot$mol$^{-1}$nm$^{-2}$ where introduced on heavy atoms. Subsequently, 50 ns of MD were performed for each structure by retaining the positional restraints on the C$\alpha$ of the protein backbone. For each trajectory, 200 frames, equally spaced in time, were extracted. Each frame was analyzed by means of SiteMap\textsuperscript{\tiny{36}} and DogSiteScorer\textsuperscript{\tiny{23}} software in order to identify druggable pockets. 
We considered as druggable a pocket falling within the following thresholds: volume $\geq$ 300 Å$^3$; depth $\geq$  10 Å; balance $\geq$  1.0; exposure $\leq$ 0.5; enclosure $\geq$  0.70; SiteScore $\geq$  0.8; DScore $\geq$  0.90; DrugScore $\geq$  0.5; SimpleScore $\geq$  0.5 (Supp. Table 1). Two pockets emerged as interesting, pocket 1, defined by residues: 127, 130, 144, 152, 159, 160, 161, 163, 167, 168, 171, 172, 173, 174, 176, 230, 237, 241, 261, 262, 264, 265, 266, 267, 268, 269, 270, 271, 272, 275, 448, 451, 452, 454, 455, 456, 459, 464, 497, 498, 499, 500, 502, 503; pocket 2, defined by residues: 236, 239, 240, 242, 243, 247, 248, 281, 282, 283, 284, 285, 286, 436, 440, 443, 591, 592, 593, 594, 596, 597, 600. The corresponding protein conformations were submitted to the virtual screening workflow (Figure 7A).

\subsection*{Preparation of the Virtual Library}
The FDA-approved drug library included compounds from the following chemical collections: Selleck (accession date 21/03/2020, 2684 compounds), Prestwick (accession date 21/03/2020, 1520 compounds) and eDrug-3D (accession date 21/03/2020, 1930 compounds). Compounds from the DrugBank collection (DrugBank Release Version 5.1.5; 1784 molecules), which includes approved, experimental and investigational drugs, were added. The four chemical libraries were merged using an in-house developed KNIME workflow to obtain a unique library of non-redundant entries (total of 9187 compounds). The final collection was prepared with the LigPrep tool\textsuperscript{\tiny{37}}. Ionization/tautomeric states were generated at pH range 7-8 using Epik\textsuperscript{\tiny{38}}. Furthermore, at most 32 stereoisomers per ligand and three lowest energy conformations per ligand ring were produced. Where not defined, all the chiral form of each stereocenter was produced. In total, 12771 docking clients were generated.

\subsection*{Virtual Screening Workflow}
For the Glide docking\textsuperscript{\tiny{22}}, the N- and C- terminal residues of the ACE2 intermediate were capped with acetyl (ACE) and N-methylamine (NMA) groups, respectively, using the Schrödinger Protein Preparation Wizard. The capped protein structure was used to generate the receptor grid, with no scaling of Van der Waals radii for non-polar receptor atoms. The docking space was centered on the centroid of the residues defining the pockets (x: 75.4, y: 58.2, z: 67.6 for pocket 1; x: 50.8, y: 57.0, z: 69.1 for pocket 2). The docking space was defined as a 35 Å$^3$ cubic box, while the diameter midpoint of docked ligands was required to remain within a smaller, nested 15 Å$^3$ cubic box. Docking experiments were performed in the Glide standard precision mode using a 0.8 factor to scale the Van der Waals radii of the ligand atoms with partial atomic charge less than 0.15. 
For the BioSolveIT docking, LeadIT (version 3.2.0) was used for protein preparation and docking parameters definition. The binding site was defined on the basis of the residues composing the identified druggable pocket. The residue protonation states, as well as the tautomeric forms, were automatically assessed in LeadIT using the ProToss method, that generates the most probable hydrogen positions on the basis of an optimal hydrogen bonding network using an empirical scoring function. The virtual screening workflow was developed by using the KNIME analytic platform and the BioSolveIT KNIME nodes. Specifically, the workflow was organized as follows: (i) the Compute LeadIT Docking node was selected to perform the docking simulations of the $\sim$11$\cdot$10$^3$ docking clients by using the FlexX algorithm\textsuperscript{\tiny{39}}. Ten poses for each ligand were produced; (ii) the Assess Affinity with HYDE in SeeSAR node generated refined binding free energy and HYDE\textsuperscript{\tiny{40}} predicted activity (HYDE\textsubscript{aff}) for each ligand pose using the HYDE rescoring function; and (iii) for each ligand, the pose with the lowest HYDE\textsubscript{aff} was extracted. 
For AutoDock docking, ligands and receptor structures were converted to the AD4 format using AutoDockTools and the Gasteiger-Marsili empirical atomic partial charges were assigned. The dimensions of the grid were 60 X 60 X 60 points, with grid points separated by a 0.375 Å. The grid was centered on the centroids of pockets 1 and 2. The Lamarckian genetic algorithm was used, and for each compound, the docking simulation was composed of 100 runs. Clustering of docked conformations was performed on the basis of their RMSD (tolerance = 2.0 Å) and the results were ranked based on the estimated free energy of binding. The obtained poses were filtered based on the Glide docking score (Glide\textsubscript{ds}), the HYDE\textsubscript{aff} and AutoDock ligand binding energy (AD\textsubscript{LBE}) as well as on the number of clusters (AD\textsubscript{NiC}). The consensus virtual screening workflow was applied for both pocket 1 and pocket 2 (Figure 7A). The best scored structures for the three software were visually inspected and selected based on their binding mode and predicted interactions. In addition, the top scoring compounds for Glide docking and SeeSAR rescoring\textsuperscript{\tiny{41}} were also evaluated. In particular, the Glide set was generated following these criteria: Glide\textsubscript{ds} $\leq$ -9 kcal/mol, giving rise to 97 molecules for pocket 1 and 30 for pocket 2.  The SeeSAR set was instead defined  by applying the following filters: (i) HYDE\textsubscript{aff} $\leq$ 5 $\mu$M; (ii) Ligand efficiency (LE) and lipophilic ligand efficiency (LLE) category from 1 to 4 (where 1 corresponds to favorable LE/LLE, 5 corresponds to unfavorable LE/LLE); (iii) Good ligand conformer quality, judged on the basis of torsional quality of the rescored pose rotamers and intramolecular clashes; (iv) no protein-ligand intermolecular clashes, giving rise to 89 molecules for pocket 1 and 205 molecules for pocket 2.

\subsection*{Cell Cultures}
HEK293, HepG2, Huh-7 and Vero cells were cultured in Dulbecco’s Minimal Essential Medium (Euroclone \#ECB7501L) containing 10\% heat-inactivated fetal bovine serum ($\Delta$56-FBS, Gibco \#10270), Penicillin/Streptomycin (Corning \#20-002-Cl), non-essential amino acids (Euroclone \#ECB3054D) and L-Glutamine (Gibco \#25030-024). Cells were passaged in 100 mm$^2$ Petri and split every 3-4 days. Cells employed in this study have not been passaged more than 20 times from the original stock.

\subsection*{Compound and Treatments}
A total 35 different compounds were tested in Vero cells for their ability to lower the expression of ACE2 (see Supp. Table 3). Each molecule, received as powder, was resuspended at 50-30 or 15 mM in DMSO (Euroclone \#APA36720250) or in Milli-Q water or Methanol. Stock solutions were prepared at 0.3, 3 and 30 mM (1000X).  To treat cells, a 1 \si{\micro}L aliquot of each stock solution was added to a well from a 24-well plate containing 1 mL of antibiotic-free cell medium (corresponding to final concentrations of 0.3, 3 and 30 \si{\micro}M). Vehicle controls were obtained by adding equivalent volumes of DMSO or Milli-Q water.

\subsection*{Cell Viability Assay (MTT)}
Cells were seeded on 24-well plates at approximately $\sim$60\% confluence. Compounds at different concentrations, or vehicle control (DMSO/MilliQ water, volume equivalent), were added after 24 h. Medium was replaced the second day, and then removed after a total of 48 h treatment. Cells were incubated with 5 mg/mL of 3-(4,5-dimethylthiazol-2-yl)-2,5-diphenyltetrazolium bromide (MTT) (Sigma \#M5655-1G) in PBS for 15 min at 37 $^{\circ}$C. After carefully removing MTT, cells were resuspended in 100 \si{\micro}L DMSO, and cell viability values obtained by a plate spectrophotometer, measuring absorbance at 560 nm.

\subsection*{Western Blotting \& Antibodies}
Cells were plated on 24-well plates at approximately $\sim$60\% confluence. Compounds at different concentrations or vehicle control (DMSO, volume equivalent) were added after 24 h, medium was replaced the second day, and then removed after a total of 48 h treatment. Samples were lysed in lysis buffer (Tris 10mM, pH 7.4, 0.5\% NP-40, 0.5\% TX-100, 150mM NaCl plus complete EDTA-free Protease Inhibitor Cocktail Tablets). Quantification of the protein was performed by BCA kit (Thermo Fisher \#23225). A 40-\si{\micro}g aliquot of total proteins were diluted 2:1 in 4X Laemmli sample buffer containing 100 mM Dithiothreitol, boiled 8 min at 95 $^{\circ}$C and loaded on SDS-PAGE, using 12\% acrylamide pre-cast gels and then transferred to hydrophilic polyvinylidene fluoride (PVDF) membranes. Membranes were blocked for 20 min in 5\% (w/v) non-fat dry milk in Tris-buffered saline containing 0.01\% Tween-20 (TBS-T). Blots were probed with anti-ACE2 antibody (1:200) (AC18Z, Santa Cruz) in 3\% (w/v) BSA in TBS-T overnight at 4 $^{\circ}$C, washed 3 times with TBS-T 10 min each, then probed with a 1:3000 dilution of horseradish conjugated goat anti-mouse for 1 h at RT. After 2 washes with TBS-T and one with Milli-Q water, signals were revealed using the ECL Select Western Blotting Detection Reagent (RPN2235, GE Healthcare) and visualized with a ChemiDoc XRS Touch Imaging System (Bio-Rad).

\subsection*{Statistical Analyses}
Statistical analyses, performed with the Prism software version 8.0 (GraphPad), included all the data points obtained, with the exception of experiments in which negative and/or positive controls did not give the expected outcome. No test for outliers was employed. Results were expressed as the mean $\pm$ standard errors. All the data were analyzed with the one-way ANOVA test, including an assessment of the normality of data, and corrected by the Dunnet post-hoc test. Probability (p) values < 0.05 were considered as significant (*). Inhibitory concentration at 50\% (IC\textsubscript{50}) or lethal dose at 50\% (LD50) values were obtained by fitting dose-response curves to a sigmoidal function using a 4PL non-linear regression model.

\subsection*{Preparation of Viral Vectors}
Retroviral particles exposing the SARS-CoV-2 spike protein were produced as follow. HEK293-T cells were seeded into 10 cm plates with selection medium (DMEM with G418 0,5 mg/mL). Once cells reached $\sim$80\% confluence, medium was replaced with DMEM containing 2,5\% FBS. Cells were then transfected with three different plasmids (MLV transfer vector, pc NCG; packaging vector, pc Gag-Pol, and env-encoding vector, pc SARS-CoV-2 spike $\Delta$C; see schematic below). The control retroviral particles was obtained by transfecting the cells with the packaging and the transfer vectors only. 

\begin{figure}[!h]
\makeatletter
\renewcommand{\fnum@figure}{\small{}}
\makeatother
\centering
\includegraphics[scale=0.45]{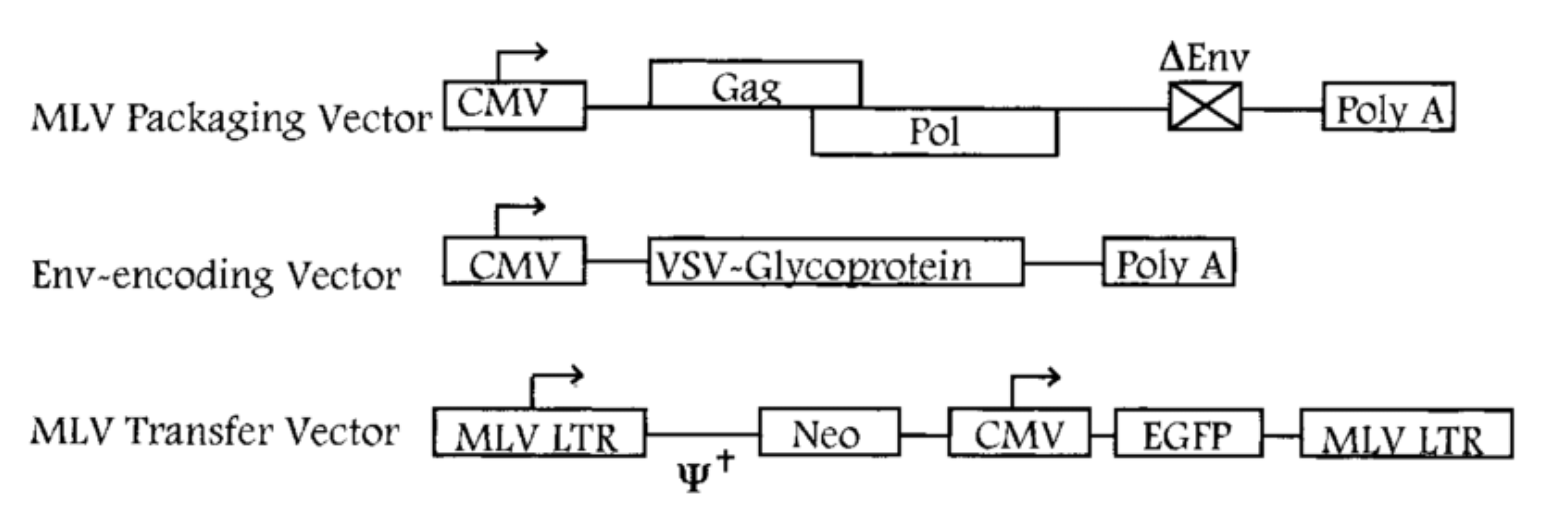}
\label{Fig9}
\end{figure}

\noindent Supernantants were collected and centrifuged at 2000 xg for 5 min, then filtered using a 0.45 \si{\micro}m filter and ultracentrifuged at 20.000 xg for 2 h Pellets were resuspended in 1X PBS and stored at -80 $^{\circ}$C.

\subsection*{Transduction Assay}
On day 1, untransfected Vero cells or HEK293 cells stably expressing the human ACE2 were seeded on 96-well plates in DMEM medium (10\% FBS, Pen/Strep, L-glutamine, non-essential amino acids). After 24 and 48 h, the medium was replaced with fresh medium containing each compound to be tested at the desired concentration. On day 4, a 3 \si{\micro}l aliquot of retroviral vectors exposing the SARS-CoV-2 spike protein, or control vectors, was added to each well. For Vero cells, this step was repeated a second time on day 5 in order to increase the number of transduced cells. Three days after transduction, cells showing the GFP fluorescence were detected with an EnSight\textsuperscript{TM} Multimode Microplate Reader.

\subsection*{Testing antiviral activity against live virus SARS-CoV-2 (MEX-BC2/2020)}
Here we report the experimental procedure for antiviral and cytotoxic assays performed by Retrovirox Inc, San Diego, California. 
\vspace{-2mm}
\subsubsection*{Sars-CoV-2 antiviral activity}
Vero E6 cells were maintained in DMEM with 10\% fetal bovine serum (FBS), hereby called DMEM10. Cells were seeded and incubated for 24 hours before being pre-incubated with test-items. Twenty-four hours post cell seeding, test samples were submitted to serial dilutions with DMEM with 2\% FBS (DMEM2) in a different plate, cell culture was removed from cells, and serial dilutions of test-items were added to the cells and incubated for 4 h at 37$^{\circ}$C in a humidified incubator. After the pre-incubation of test-items with target cells, cells were challenged with the viral inoculum resuspended in DMEM2. The amount of viral inoculum was previously titrated to result in a linear response inhibited by antivirals with activity against SARS-CoV-2. Cell culture media with the virus inoculum was not removed, and the test-items and virus were maintained in the media for the duration of the assay (96 h). After this period the extent of cell viability was monitored with the neutral red (NR) uptake assay. The virus-induced CPE was monitored under the microscope after 3 days of infection and at day 4 cells were stained with neutral red to monitor cell viability. Viable cells incorporate neutral red in their lysosomes. The uptake relies on the ability of live cells to maintain the pH inside the lysosomes lower than in the cytoplasm. This process requires ATP. Inside the lysosome the dye becomes charged and is retained. After a 3 h incubation with neutral red (0.033\%), the extra dye was washed and the neutral red taken by lysosomes was then extracted for 15 minutes with a solution containing 50\% ethanol and 1\% acetic acid to monitor absorbance at 540 nm. Ziprasidone, Buclizine and Beclabuvir were evaluated in quadruplicates (duplicates in two separate plates), while Artefenomel was evaluated in double triplicated (triplicates in two separate plates) using serial 3-fold dilutions. Controls included uninfected cells (“mock-infected”), and infected cells to which only vehicle was added. Some cells were treated with GS-441524. GS-441524 is the main metabolite of remdesivir, a broad spectrum antiviral that blocks the RNA polymerase of SARS-CoV-2. The average absorbance at 540 nm (A540) observed in infected cells (in the presence of vehicle alone) was calculated, and then subtracted from all samples to determine the inhibition of the virus induced CPE. Data points were then normalized to the average A540 signal observed in uninfected cells (“mock”) after subtraction of the absorbance signal observed in infected cells. In the neutral red CPE-based assay, uninfected cells remained viable and uptake the dye at higher levels than non-viable cells. In the absence of antiviral agents the virus-induced CPE kills infected cells and leads to lower A540 (this value equals 0\% inhibition). By contrast, incubation with the antiviral agent (GS-441524) prevents the virus induced CPE and leads absorbance levels similar to those observed in uninfected cells. Full recovery of cell viability in infected cells represent 100\% inhibition of virus replication. 
\vspace{-2mm}
\subsubsection*{Cytotoxicity assays}
Uninfected cells were incubated with eight concentrations of test-items or control inhibitors dilutions. The incubation temperature and duration of the incubation period mirrored the conditions of the prevention of virus-induced CPE assay, and cell viability was evaluated with the neutral red uptake method but this time utilizing uninfected cells, otherwise the procedure was the same as the one used for the antiviral assays. The extent of viability was monitored by measuring absorbance at 540 nm. When analyzing the data, background levels obtained from wells with no cells were subtracted from all data-points. Absorbance readout values were given as a percentage of the average signal observed in uninfected cells treated with vehicle alone. The average signal obtained in wells with no cells was subtracted from all samples. Readout values were given as a percentage of the average signal observed in uninfected cells treated with vehicle alone (DMEM2). The signal- to-background (S/B) obtained was 29-fold. DMSO was used as a cytotoxic compound control in the viability assays. DMSO blocked cell viability by more than 99\% when tested at 10\%.

\section*{Supplementary Materials}
\noindent Atomic resolution structures of the folding intermediates together with the corresponding druggable pockets are freely available upon request.

\section*{Acknowledgements}
The computational reconstruction of ACE2 folding pathway was performed thanks to the strong support of the high performance computational infrastructures made available by Italian National Institute of Nuclear Physics (INFN). This work was supported by a grant from Fondazione Telethon (Italy, TCP14009). GS is a recipient of a fellowship from Fondazione Telethon. EB is an Assistant Telethon Scientist at the Dulbecco Telethon Institute.

\section*{Author Contribution}
Conceived and designed the analyses: all authors \\
Performed the analyses of folding simulations: AB, LT, GS\\
Performed the virtual screening: AA\\
Performed cellular tests: TM, VB, FP\\
Analyzed the data: AB, LT, GS, AA, MLB and PF\\
Wrote the paper: GS, PF, AA, MLB and EB \\
All authors edited the manuscript

\section*{Competing Interests}
TM, AB, LT, AA and LP have direct involvement in the ongoing research at Sibylla Biotech SRL, a spin-off company of University of Trento, University of Perugia and INFN (National Institute for Nuclear Physics). The company exploits the PPI-FIT technology for drug discovery in a wide variety of human pathologies, except prion diseases. GS, GL, MLB, EB, PF, LP are co-founders and shareholders of the company (www.sibyllabiotech.it).

\newpage 

\section*{References}

\begin{enumerate}
	\item Velavan TP, Meyer CG. The COVID-19 epidemic. Trop Med Int Health. 2020;25(3):278-80. Epub 2020/02/14. doi: 10.1111/tmi.13383. PubMed PMID: 32052514
	\item Rothan HA, Byrareddy SN. The epidemiology and pathogenesis of coronavirus disease (COVID-19) outbreak. J Autoimmun. 2020;109:102433. Epub 2020/03/03. doi: 10.1016/j.jaut.2020.102433. PubMed PMID: 32113704; PubMed Central PMCID: PMCPMC7127067.
	\item Li Q, Guan X, Wu P, Wang X, Zhou L, Tong Y, et al. Early Transmission Dynamics in Wuhan, China, of Novel Coronavirus-Infected Pneumonia. N Engl J Med. 2020;382(13):1199-207. Epub 2020/01/30. doi: 10.1056/NEJMoa2001316. PubMed PMID: 31995857; PubMed Central PMCID: PMCPMC7121484.
	\item Lim YX, Ng YL, Tam JP, Liu DX. Human Coronaviruses: A Review of Virus-Host Interactions. Diseases. 2016;4(3). Epub 2016/07/25. doi: 10.3390/diseases4030026. PubMed PMID: 28933406; PubMed Central PMCID: PMCPMC5456285.
	\item Wu F, Zhao S, Yu B, Chen YM, Wang W, Song ZG, et al. A new coronavirus associated with human respiratory disease in China. Nature. 2020;579(7798):265-9. Epub 2020/02/06. doi: 10.1038/s41586-020-2008-3. PubMed PMID: 32015508; PubMed Central PMCID: PMCPMC7094943.
	\item Tai W, He L, Zhang X, Pu J, Voronin D, Jiang S, et al. Characterization of the receptor-binding domain (RBD) of 2019 novel coronavirus: implication for development of RBD protein as a viral attachment inhibitor and vaccine. Cell Mol Immunol. 2020. Epub 2020/03/24. doi: 10.1038/s41423-020-0400-4. PubMed PMID: 32203189.
	\item Shang J, Ye G, Shi K, Wan Y, Luo C, Aihara H, et al. Structural basis of receptor recognition by SARS-CoV-2. Nature. 2020. Epub 2020/04/01. doi: 10.1038/s41586-020-2179-y. PubMed PMID: 32225175.
	\item Lan J, Ge J, Yu J, Shan S, Zhou H, Fan S, et al. Structure of the SARS-CoV-2 spike receptor-binding domain bound to the ACE2 receptor. Nature. 2020. Epub 2020/04/01. \\doi: 10.1038/s41586-020-2180-5. PubMed PMID: 32225176.
	\item Hoffmann M, Kleine-Weber H, Schroeder S, Kruger N, Herrler T, Erichsen S, et al. SARS-CoV-2 Cell Entry Depends on ACE2 and TMPRSS2 and Is Blocked by a Clinically Proven Protease Inhibitor. Cell. 2020. Epub 2020/03/07. doi: 10.1016/j.cell.2020.02.052. PubMed PMID: 32142651; PubMed Central PMCID: PMCPMC7102627.
	\item Burrell LM, Johnston CI, Tikellis C, Cooper ME. ACE2, a new regulator of the renin-angiotensin system. Trends Endocrinol Metab. 2004;15(4):166-9. Epub 2004/04/28. \\ doi: 10.1016/j.tem.2004.03.001. PubMed PMID: 15109615; PubMed Central \\ PMCID: PMCPMC7128798.
	\item Wiener RS, Cao YX, Hinds A, Ramirez MI, Williams MC. Angiotensin converting enzyme 2 is primarily epithelial and is developmentally regulated in the mouse lung. J Cell Biochem. 2007;101(5):1278-91. Epub 2007/03/07. doi: 10.1002/jcb.21248. PubMed PMID: 17340620.
	\item Lukassen S, Chua RL, Trefzer T, Kahn NC, Schneider MA, Muley T, et al. SARS-CoV-2 receptor ACE2 and TMPRSS2 are primarily expressed in bronchial transient secretory cells. EMBO J. 2020:e105114. Epub 2020/04/15. doi: 10.15252/embj.2020105114. PubMed PMID: 32285974.
	\item Turner AJ, Hooper NM. The angiotensin-converting enzyme gene family: genomics and pharmacology. Trends Pharmacol Sci. 2002;23(4):177-83. Epub 2002/04/05. doi: 10.1016/s0165-6147(00)01994-5. PubMed PMID: 11931993.
	\item Tikellis C, Bernardi S, Burns WC. Angiotensin-converting enzyme 2 is a key modulator of the renin-angiotensin system in cardiovascular and renal disease. Curr Opin Nephrol Hypertens. 2011;20(1):62-8. Epub 2010/11/26. doi: 10.1097/MNH.0b013e328341164a. PubMed PMID: 21099686.
	\item Iwata M, Cowling RT, Yeo SJ, Greenberg B. Targeting the ACE2-Ang-(1-7) pathway in cardiac fibroblasts to treat cardiac remodeling and heart failure. J Mol Cell Cardiol. 2011;51(4):542-7. Epub 2010/12/15. doi: 10.1016/j.yjmcc.2010.12.003. PubMed PMID: 21147120; PubMed Central PMCID: PMCPMC3085048.
	\item Tu YF, Chien CS, Yarmishyn AA, Lin YY, Luo YH, Lin YT, et al. A Review of SARS-CoV-2 and the Ongoing Clinical Trials. Int J Mol Sci. 2020;21(7). Epub 2020/04/16. doi: 10.3390/ijms21072657. PubMed PMID: 32290293.
	\item Zhang H, Penninger JM, Li Y, Zhong N, Slutsky AS. Angiotensin-converting enzyme 2 (ACE2) as a SARS-CoV-2 receptor: molecular mechanisms and potential therapeutic target. Intensive Care Med. 2020;46(4):586-90. Epub 2020/03/04. doi: 10.1007/s00134-020-05985-9. PubMed PMID: 32125455; PubMed Central PMCID: PMCPMC7079879.
	\item Giovanni Spagnolli TM, Andrea Astolfi, Silvia Biggi, Paolo Brunelli, Michela, Libergoli AI, Simone Orioli, Alberto Boldrini, Luca Terruzzi, Giulia Maietta, Marta, Rigoli NLL, Leticia C. Fernandez, Laura Tosatto, Luise Linsenmeier, Beatrice, Vignoli GP, Dino Gasparotto, Maria Pennuto, Graziano Guella, Marco Canossa,, Hermann Clemens Altmeppen GL, Stefano Biressi, Manuel Martin Pastor, Jesús R., Requena IM, Maria Letizia Barreca, Pietro Faccioli and Emiliano Biasini. Pharmacological Protein Inactivation by Targeting Folding Intermediates. BioRXiv. 2020. doi: https://doi.org/10.1101/2020.03.31.018069.
	\item Spagnolli G, Rigoli M, Orioli S, Sevillano AM, Faccioli P, Wille H, et al. Full atomistic model of prion structure and conversion. PLoS Pathog. 2019;15(7):e1007864. Epub 2019/07/12. doi: 10.1371/journal.ppat.1007864. PubMed PMID: 31295325; \\ PubMed Central PMCID: PMCPMC6622554.
	\item A Beccara S, Fant L, Faccioli P. Variational scheme to compute protein reaction pathways using atomistic force fields with explicit solvent. Phys Rev Lett. 2015;114(9):098103. Epub 2015/03/21. doi: 10.1103/PhysRevLett.114.098103. PubMed PMID: 25793854.
	\item Huang J, Rauscher S, Nawrocki G, Ran T, Feig M, de Groot BL, et al. CHARMM36m: an improved force field for folded and intrinsically disordered proteins. Nat Methods. 2017;14(1):71-3. Epub 2016/11/08. doi: 10.1038/nmeth.4067. PubMed PMID: 27819658; PubMed Central PMCID: PMCPMC5199616
	\item Schrödinger L, New York, NY, 2019. Schrödinger Release 2019-3: SiteMap.
	\item Volkamer A, Kuhn D, Rippmann F, Rarey M. DoGSiteScorer: a web server for automatic binding site prediction, analysis and druggability assessment. Bioinformatics. 2012;28(15):2074-5. Epub 2012/05/26. doi: 10.1093/bioinformatics/bts310. PubMed PMID: 22628523.
	\item GmbH B. LeadIT version 3.2.0. wwwbiosolveitde/LeadIT.
	\item Morris GM, Huey R, Lindstrom W, Sanner MF, Belew RK, Goodsell DS, et al. AutoDock4 and AutoDockTools4: Automated docking with selective receptor flexibility. J Comput Chem. 2009;30(16):2785-91. Epub 2009/04/29. doi: 10.1002/jcc.21256. PubMed PMID: 19399780; PubMed Central PMCID: PMCPMC2760638.
		\item 	Sangeun Jeon MK, Jihye Lee, Inhee Choi, Soo Young Byun, Soonju Park, David Shum, Seungtaek Kim. Identification of antiviral drug candidates against SARS-CoV-2 from FDA-approved drugs. biorxiv. 2020. doi: https://doi.org/10.1101/2020.03.20.999730
	\item Liu J, Cao R, Xu M, Wang X, Zhang H, Hu H, et al. Hydroxychloroquine, a less toxic derivative of chloroquine, is effective in inhibiting SARS-CoV-2 infection in vitro. Cell Discov. 2020;6:16. Epub 2020/03/21. doi: 10.1038/s41421-020-0156-0. PubMed PMID: 32194981; PubMed Central PMCID: PMCPMC7078228.
	\item Devaux CA, Rolain JM, Colson P, Raoult D. New insights on the antiviral effects of chloroquine against coronavirus: what to expect for COVID-19? Int J Antimicrob Agents. 2020:105938. Epub 2020/03/17. doi: 10.1016/j.ijantimicag.2020.105938. PubMed PMID: 32171740; PubMed Central PMCID: PMCPMC7118659.
	\item Caillaud M, El Madani M, Massaad-Massade L. Small interfering RNA from the lab discovery to patients' recovery. J Control Release. 2020;321:616-28. Epub 2020/02/23. \\ doi: 10.1016/j.jconrel.2020.02.032. PubMed PMID: 32087301.
	\item Wang D, Zhang F, Gao G. CRISPR-Based Therapeutic Genome Editing: Strategies and In Vivo Delivery by AAV Vectors. Cell. 2020;181(1):136-50. Epub 2020/04/04. \\ doi: 10.1016/j.cell.2020.03.023. PubMed PMID: 32243786.
	\item Toure M, Crews CM. Small-Molecule PROTACS: New Approaches to Protein Degradation. Angew Chem Int Ed Engl. 2016;55(6):1966-73. Epub 2016/01/13. doi: 10.1002/anie.201507978. PubMed PMID: 26756721.

	\item Van Der Spoel D, Lindahl E, Hess B, Groenhof G, Mark AE, Berendsen HJ. GROMACS: fast, flexible, and free. J Comput Chem. 2005;26(16):1701-18. Epub 2005/10/08. doi: 10.1002/jcc.20291. PubMed PMID: 16211538.
	\item Paci E, Karplus M. Forced unfolding of fibronectin type 3 modules: an analysis by biased molecular dynamics simulations. J Mol Biol. 1999;288(3):441-59. Epub 1999/05/18. \\ doi: 10.1006/jmbi.1999.2670. PubMed PMID: 10329153.
	\item Camilloni C, Broglia RA, Tiana G. Hierarchy of folding and unfolding events of protein G, CI2, and ACBP from explicit-solvent simulations. J Chem Phys. 2011;134(4):045105. Epub 2011/02/02. doi: 10.1063/1.3523345. PubMed PMID: 21280806.
	\item Bussi G, Donadio D, Parrinello M. Canonical sampling through velocity rescaling. J Chem Phys. 2007;126(1):014101. Epub 2007/01/11. doi: 10.1063/1.2408420. PubMed PMID: 17212484.
	\item Halgren TA. Identifying and characterizing binding sites and assessing druggability. J Chem Inf Model. 2009;49(2):377-89. Epub 2009/05/13. doi: 10.1021/ci800324m. PubMed PMID: 19434839.
	\item Mahto MK, Yellapu NK, Kilaru RB, Chamarthi NR, Bhaskar M. Molecular designing and in silico evaluation of darunavir derivatives as anticancer agents. Bioinformation. 2014;10(4):221-6. Epub 2014/06/27. doi: 10.6026/97320630010221. PubMed PMID: 24966524; PubMed Central PMCID: PMCPMC4070053.
	\item Shelley JC, Cholleti A, Frye LL, Greenwood JR, Timlin MR, Uchimaya M. Epik: a software program for pK( a ) prediction and protonation state generation for drug-like molecules. J Comput Aided Mol Des. 2007;21(12):681-91. Epub 2007/09/28. doi: 10.1007/s10822-007-9133-z. PubMed PMID: 17899391.
	\item Rarey M, Kramer B, Lengauer T, Klebe G. A fast flexible docking method using an incremental construction algorithm. J Mol Biol. 1996;261(3):470-89. Epub 1996/08/23. \\ doi: 10.1006/jmbi.1996.0477. PubMed PMID: 8780787.
	\item Reulecke I, Lange G, Albrecht J, Klein R, Rarey M. Towards an integrated description of hydrogen bonding and dehydration: decreasing false positives in virtual screening with the HYDE scoring function. ChemMedChem. 2008;3(6):885-97. Epub 2008/04/04. doi: 10.1002/cmdc.200700319. PubMed PMID: 18384086.
	\item GmbH B. SeeSAR Version 9.2. wwwbiosolveitde/SeeSAR.

\end{enumerate}

\newpage

\section*{Supplementary Figures}

\begin{figure}[!h]
\makeatletter
\renewcommand{\fnum@figure}{\small{Supp. Figure 1}}
\makeatother
\centering
\includegraphics[scale=0.4]{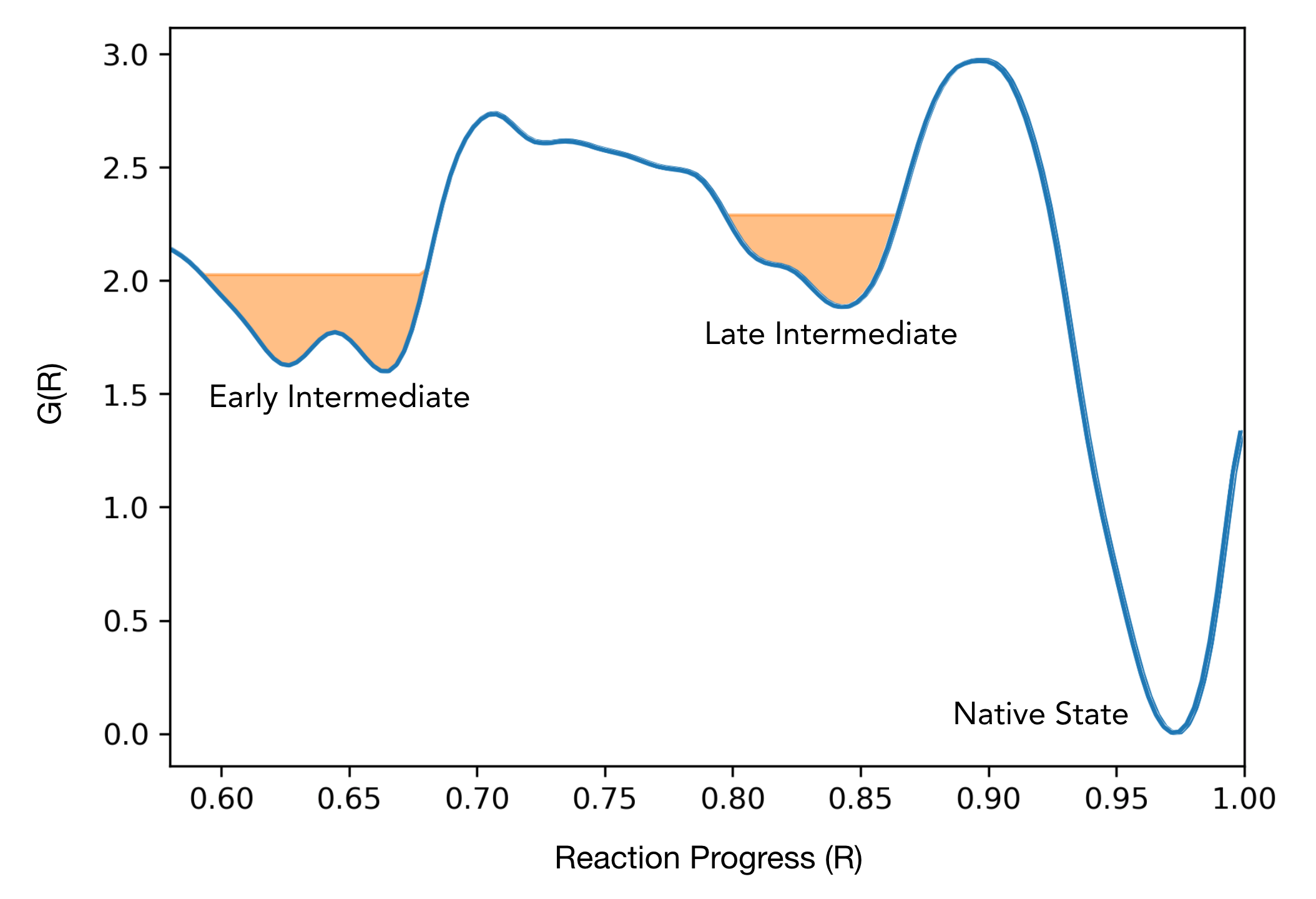}
\caption{\footnotesize{\textbf{Representation of the selected regions defined by the energy wells. }The selected regions defining the two energy wells are depicted in orange. In particular, the early-intermediate region is defined with R between 0.59 and 0.68, while the late-intermediate region is defined with R between 0.797 and 0.865.}}
\label{Sup1}
\end{figure}

\vspace{1.5cm}

\begin{figure}[!h]
\makeatletter
\renewcommand{\fnum@figure}{\small{Supp. Figure 2}}
\makeatother
\centering
\includegraphics[scale=0.37]{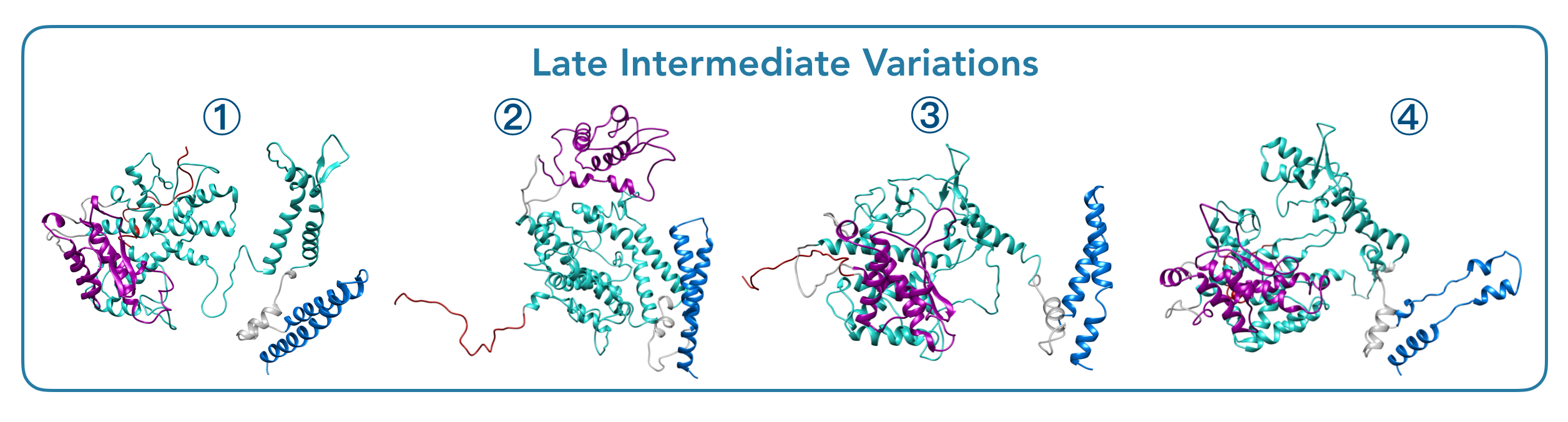}
\caption{\footnotesize{\textbf{Selected structure variants for each cluster. }Representative conformations of the additional structures selected for each cluster.}}
\label{Sup2}
\end{figure}

\clearpage

\begin{figure}[!h]
\makeatletter
\renewcommand{\fnum@figure}{\small{Supp. Figure 3}}
\makeatother
\centering
\includegraphics[scale=0.5]{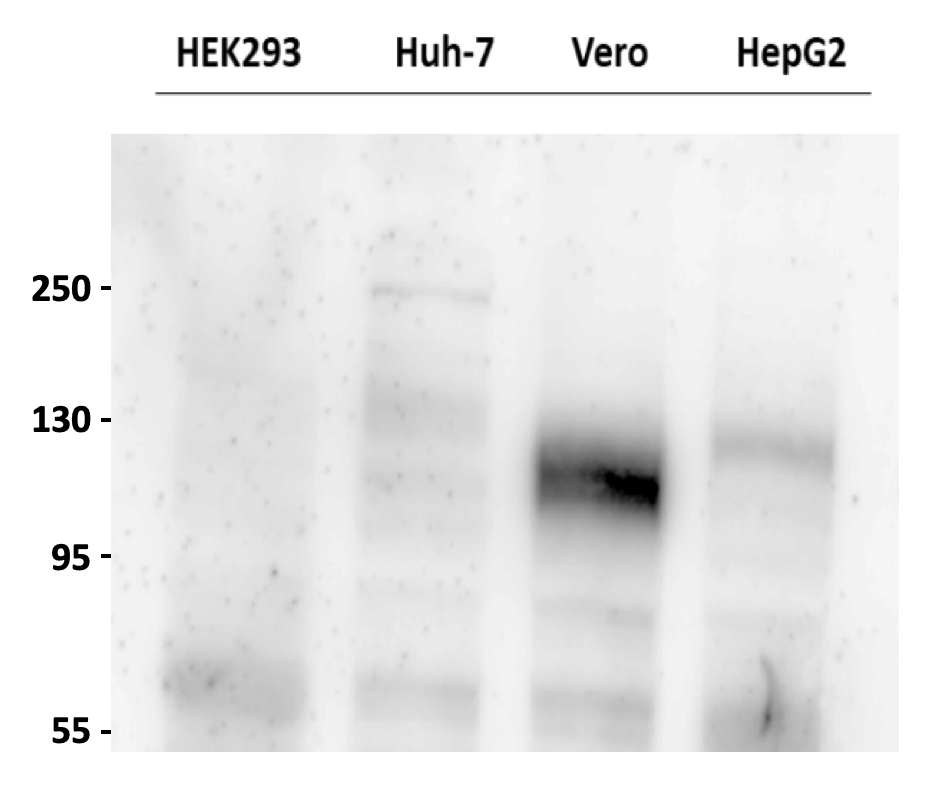}
\caption{\footnotesize{\textbf{ACE2 expression in different cell lines.} The expression of ACE2 in the indicated cell lines was detected by western blotting using an anti-ACE2 antibody (AC18Z).}}
\label{Sup3}
\end{figure}

\newpage

\begin{figure}[!h]
\makeatletter
\renewcommand{\fnum@figure}{\small{Supp. Figure 4}}
\makeatother
\centering
\includegraphics[scale=0.46]{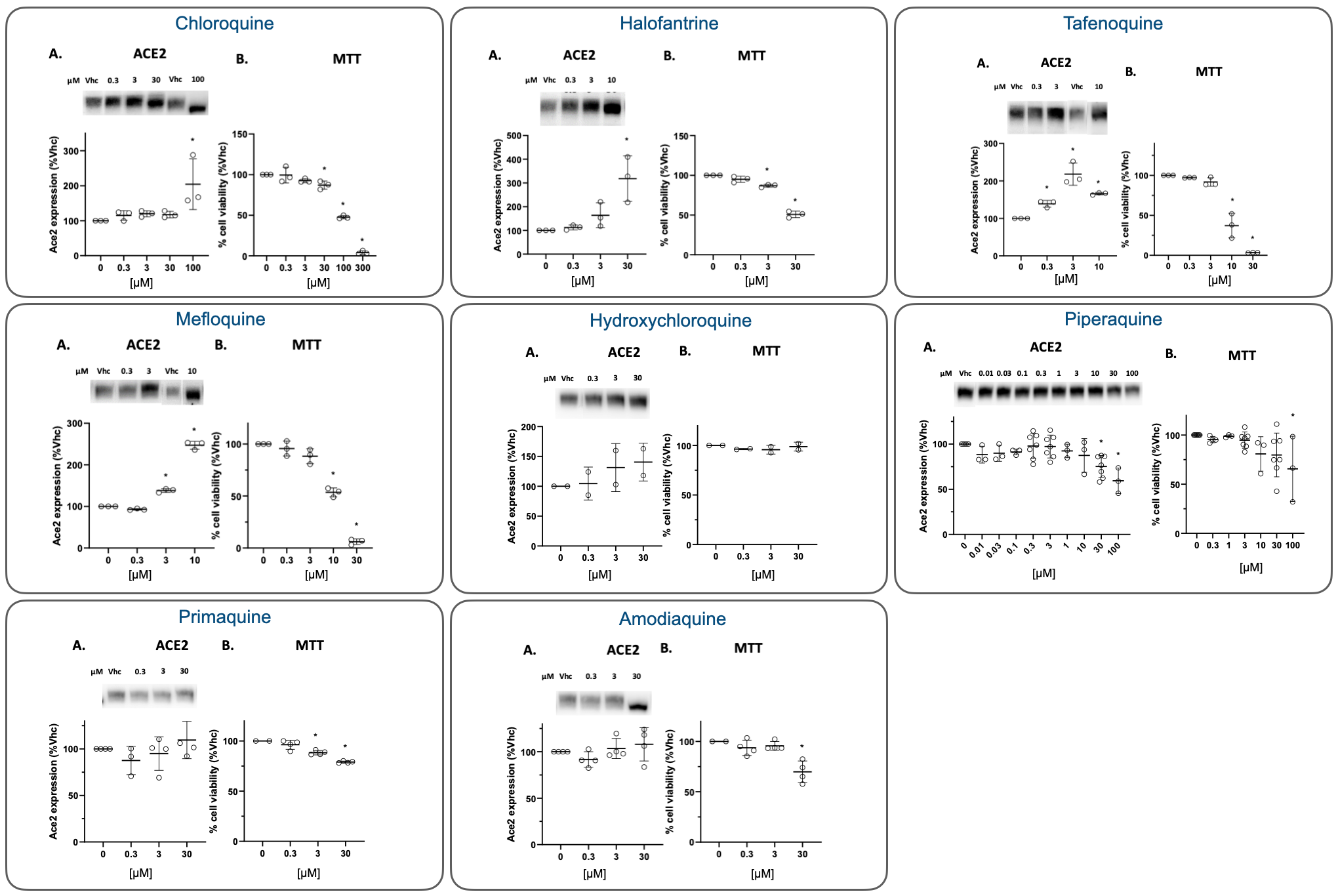}
\caption{\footnotesize{\textbf{Lack of ACE2 lowering effects for Chloroquine and analogues.} Untransfected Vero cells exposed to different concentrations of each compound or vehicle for 48 h, were lysed and analyzed by western blotting. Signals were detected by using relevant primary and secondary antibodies. Images above the graphs are representative examples of different experiments (n $\geq$ 3). The graphs show the densitometric quantification of the levels of ACE2 (A). Each signal was normalized on the corresponding total protein lane (detected by UV) and expressed as the percentage of the level in vehicle (Vhc)-treated controls. The intrinsic toxicity of each molecule was assessed by MTT assay (B). Statistically significant differences are indicated by the asterisk (* p < 0.05).}}
\label{Sup4}
\end{figure}

\newpage

\section*{Supplementary Tables}

\begin{figure}[!h]
\makeatletter
\renewcommand{\fnum@figure}{\small{Supp. Table 1}}
\makeatother
\centering
\includegraphics[scale=0.41]{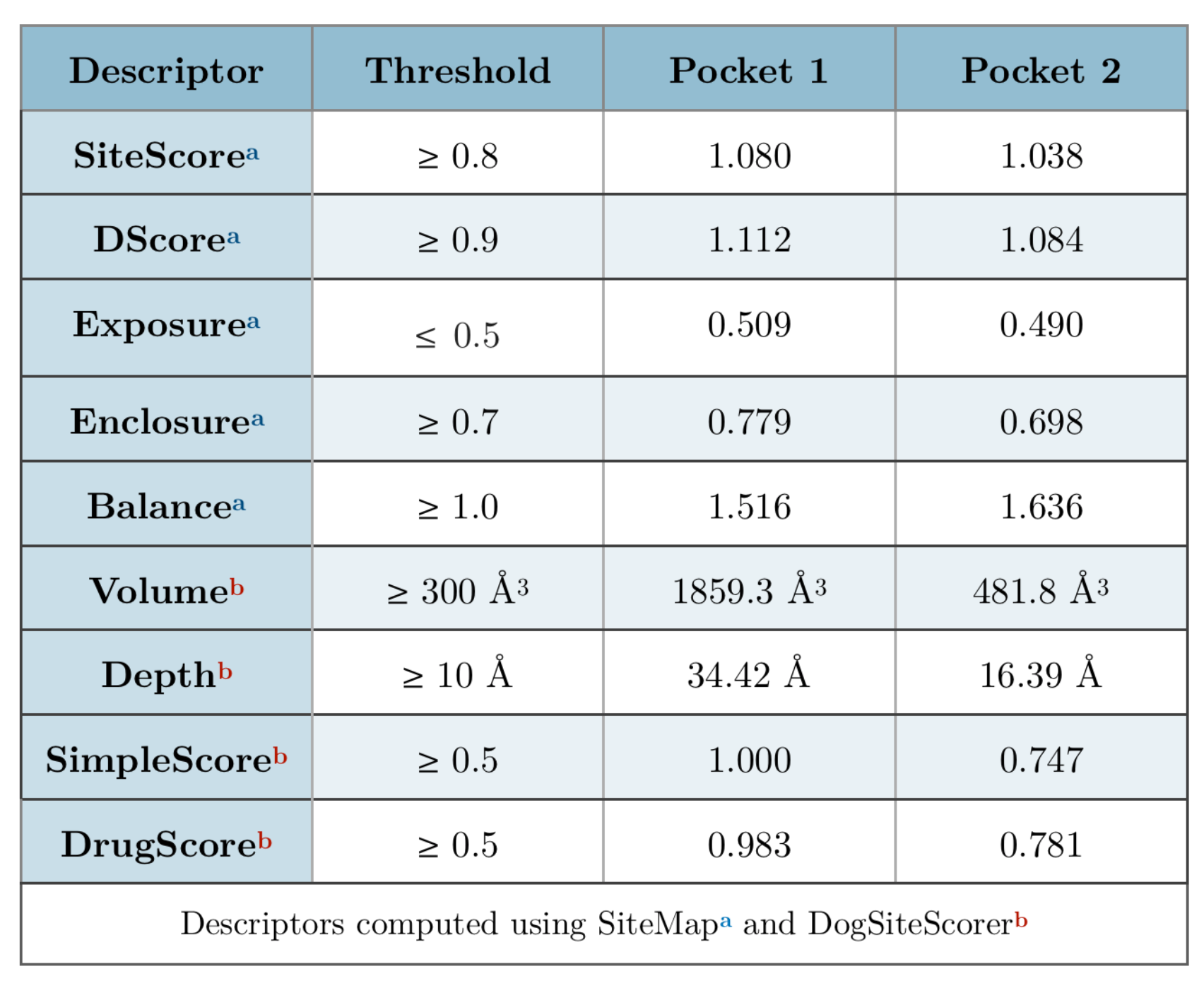}
\caption{\footnotesize{\textbf{Druggability descriptors of the identified pockets. }The table shows the values of the druggability descriptors computed using SiteMap and DogSiteScorer for the two pockets. The threshold describes the value required by a descriptor for the definition of a good druggable site.}}
\label{Tab1}
\end{figure}

\newpage

\begin{figure}[!h]
\makeatletter
\renewcommand{\fnum@figure}{\small{Supp. Table 2}}
\makeatother
\centering
\includegraphics[scale=0.82]{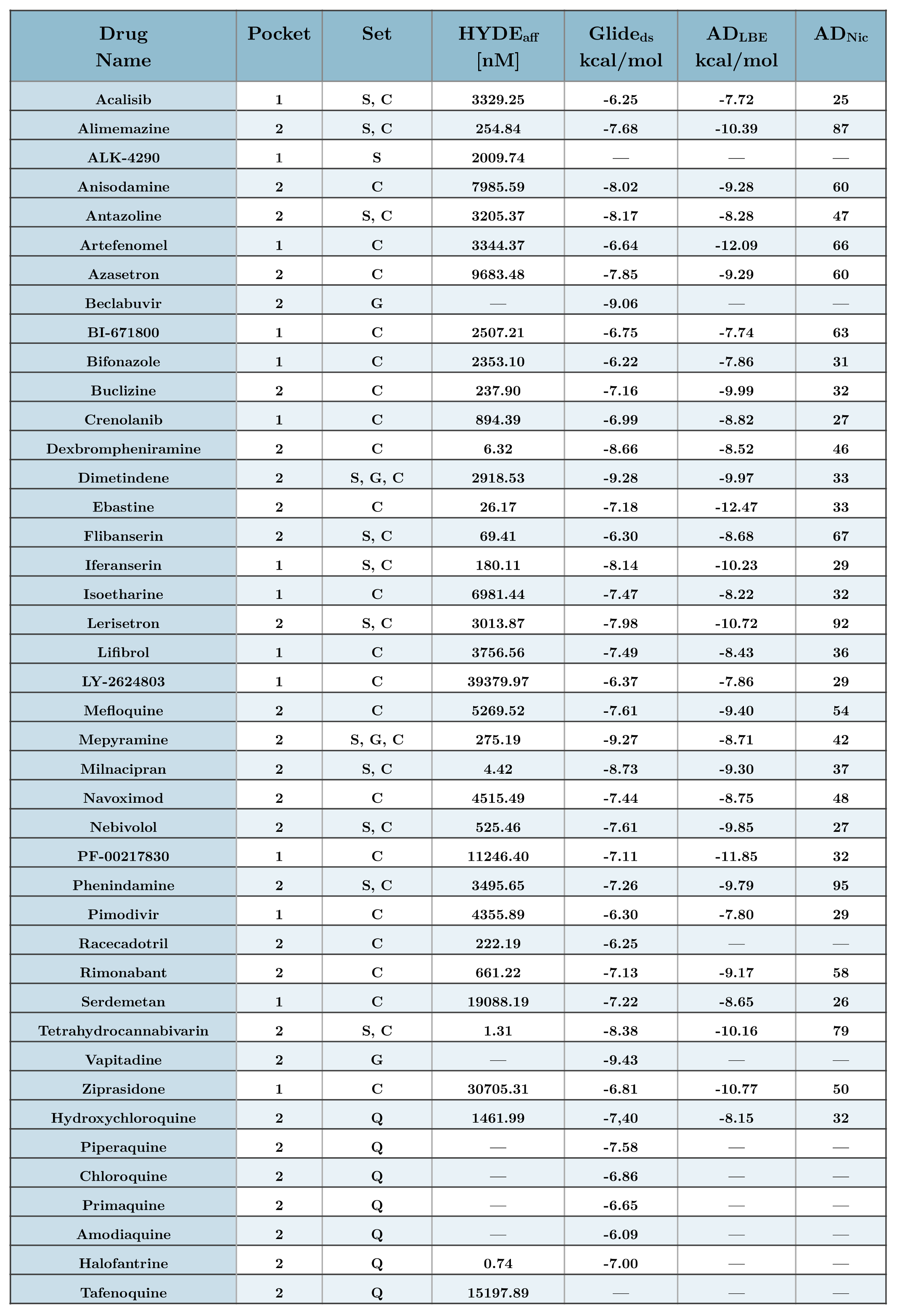}
\caption{\footnotesize{\textbf{Predicted ligands for pockets 1 and 2 of the ACE2 folding intermediate. }The table shows the different hits emerging from the virtual screening as well as the mefloquine analogues analyzed subsequently. Column Set indicates whether the molecule was identified by consensus (C), Glide (G) or SeeSAR (S) schemes; while (Q) indicates the quinoline derivatives. Affinity predictors calculated with the different software are also reported: HYDE\textsubscript{aff} (LeadIT), Glide\textsubscript{ds} (Glide), AD\textsubscript{LBE} and AD\textsubscript{NiC} (AutoDock).}}
\label{Tab2}
\end{figure}

\newpage

\begin{figure}[!h]
\makeatletter
\renewcommand{\fnum@figure}{\small{Supp. Table 3}}
\makeatother
\centering
\includegraphics[scale=0.64]{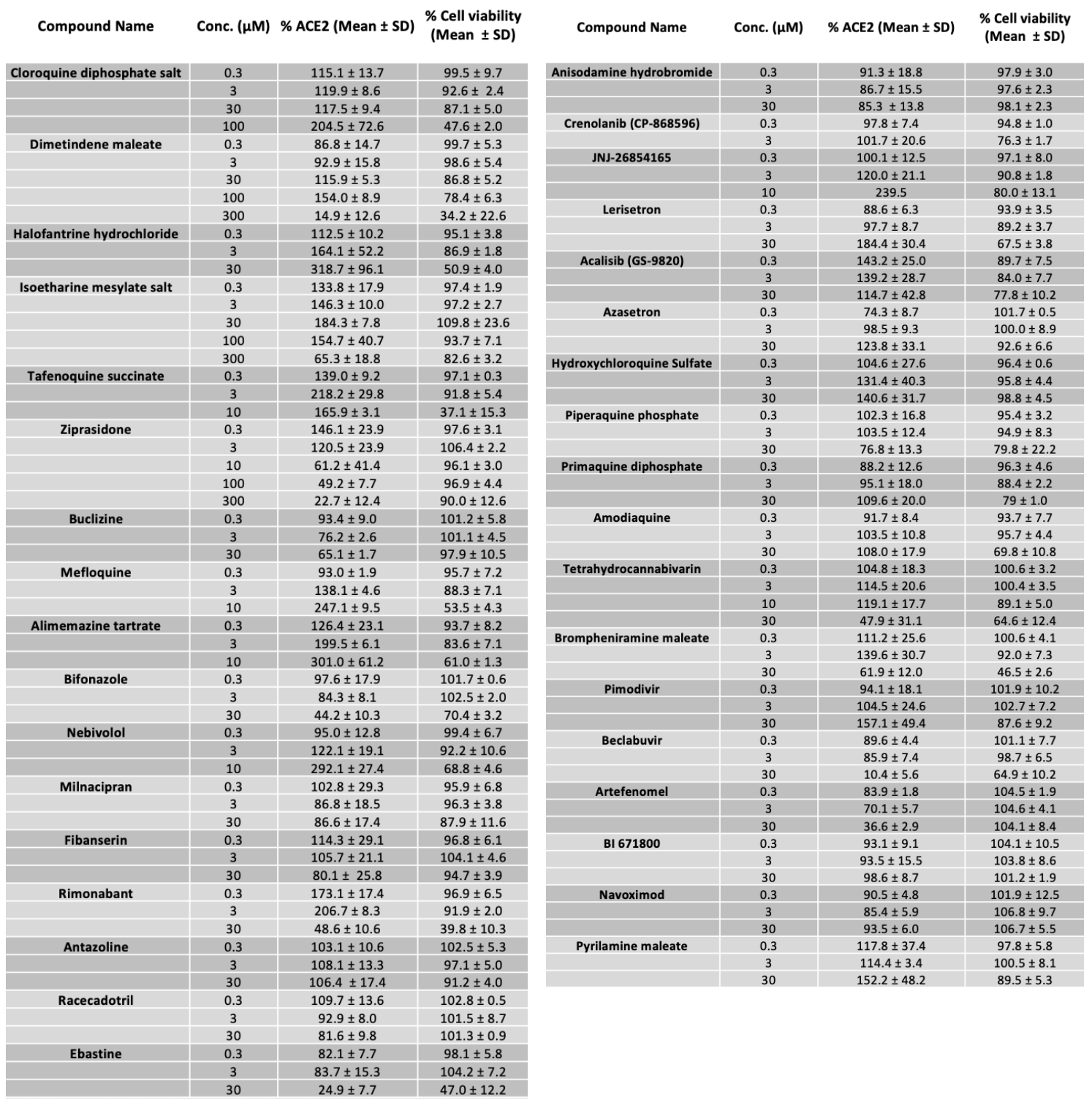}
\caption{\footnotesize{\textbf{ACE2 expression levels and cytotoxicity for the 35 compounds tested.} Individual values for ACE2 expression are listed above. Each signal was normalized on the corresponding total protein lane (detected by UV, and allowed by the enhanced tryptophan fluorescence technology of stain-free gels) and expressed as the percentage of the level in vehicle (Vhc)-treated controls. The colorimetric MTT assay was employed to monitor the intrinsic toxicity of each compound. Values represent the average of 3-6 independent experiments (depending on observed variability), and expressed as percentage of Vhc-treated controls. Employed concentrations were chosen based on the solubility of the compounds.}}
\label{Tab3}
\end{figure}

\end{document}